\documentclass{ieeeaccess}
\usepackage{cite}
\usepackage{amsmath,amssymb,amsfonts}
\usepackage{algorithmic}
\usepackage{graphicx}
\usepackage{textcomp}

\usepackage{url}
\usepackage{amsfonts}
\usepackage{booktabs}
\usepackage{multirow}
\usepackage{xspace}
\usepackage[subrefformat=parens]{subcaption}
\usepackage{cite}

\usepackage[normalem]{ulem}
\DeclareRobustCommand{\Erase}{\bgroup\markoverwith{\textcolor{red}{\rule[.5ex]{2pt}{0.4pt}}}\ULon}

\def\BibTeX{{\rm B\kern-.05em{\sc i\kern-.025em b}\kern-.08em
    T\kern-.1667em\lower.7ex\hbox{E}\kern-.125emX}}

\begin{document}
\history{Date of publication xxxx 00, 0000, date of current version xxxx 00, 0000.}
\doi{10.1109/ACCESS.2017.DOI}

\title{Comprehensive Comparisons of Uniform Quantization in Deep Image Compression}
\author{\uppercase{Koki Tsubota} \IEEEmembership{Graduate Student Member, IEEE},
\uppercase{Kiyoharu Aizawa} \IEEEmembership{Fellow, IEEE}}
\address[]{Department of Information and Communication Engineering, The University of Tokyo, Tokyo 113-8654, Japan.}
\tfootnote{This work is partially supported by Hitachi, Ltd. and Japan Society for the Promotion of Science (JSPS) KAKENHI under Grant 22J13735.}

\corresp{Corresponding author: Koki Tsubota (e-mail: tsubota@hal.t.u-tokyo.ac.jp).}

\markboth
{Tsubota \headeretal: Comprehensive Comparisons of Uniform Quantization in Deep Image Compression}
{Tsubota \headeretal: Comprehensive Comparisons of Uniform Quantization in Deep Image Compression}

\def\x{{\mathbf x}}
\def\y{{\mathbf y}}

\makeatletter
\DeclareRobustCommand\onedot{\futurelet\@let@token\@onedot}
\def\@onedot{\ifx\@let@token.\else.\null\fi\xspace}
\def\eg{\emph{e.g}\onedot} \def\Eg{\emph{E.g}\onedot}
\def\ie{\emph{i.e}\onedot} \def\Ie{\emph{I.e}\onedot}
\def\cf{\emph{c.f}\onedot} \def\Cf{\emph{C.f}\onedot}
\def\etc{\emph{etc}\onedot} \def\vs{\emph{vs}\onedot}
\def\wrt{w.r.t\onedot} \def\dof{d.o.f\onedot}
\def\etal{\emph{et al}\onedot}
\makeatother

\begin{abstract}
    In deep image compression, uniform quantization is applied to latent representations obtained by using an auto-encoder architecture for reducing bits and entropy coding.
    Quantization is a problem encountered in the end-to-end training of deep image compression.
    Quantization's gradient is zero, and it cannot backpropagate meaningful gradients.
    Many methods have been proposed to address the approximations of quantization to obtain gradients.
    However, there have not been equitable comparisons among them.
    In this study, we comprehensively compare the existing approximations of uniform quantization.
    Furthermore, we evaluate possible combinations of quantizers for the decoder and the entropy model, as the approximated quantizers can be different for them.
    We conduct experiments using three network architectures on two test datasets.
    The experimental results reveal that the best approximated quantization differs by the network architectures,
    and the best approximations of the three are different from the original ones used for the architectures.
    We also show that the combination of quantizers that uses universal quantization for the entropy model and differentiable soft quantization for the decoder is a comparatively good choice for different architectures and datasets.
\end{abstract}


\titlepgskip=-15pt

\maketitle

\section{Introduction}
\label{sec:intro}
\PARstart{I}{mage} compression is a fundamental image processing task.
It saves costs for storage and Internet traffic by reducing the bits of images.
Traditional standards such as JPEG~\cite{JPEG}, JPEG2000~\cite{JPEG2000}, BPG~\cite{BPG}, and WebP~\cite{WebP} are image compressions using hand-crafted modules, and each module is optimized separately.
Recent studies on image compression have been conducted based on deep neural networks (deep image compression)~\cite{conf/iclr/BalleLS17,conf/nips/MinnenBT18}.
Deep image compression optimizes modules in an end-to-end manner, in contrast to traditional methods.
Recently, considerable progress has been made, resulting in higher performance over traditional compression methods ~\cite{conf/cvpr/Cheng2020,guo2021causal}.

\begin{figure*}[t]
  \centering
  \includegraphics[width=\hsize]{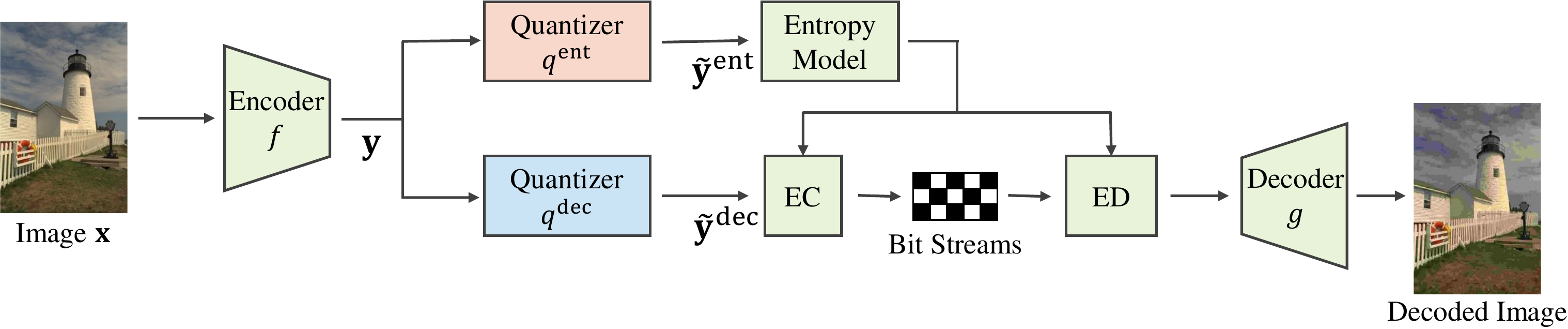}
  \caption{Outline of our method. We prepare two uniform quantizers with the same interval for a decoder and an entropy model. We then evaluate the performance when using an approximation method for each quantizer. Using the same approximation method for these two quantizers is equivalent to using the approximation method with a single quantizer in our implementation. EC and ED denote entropy coding and decoding, respectively.}
  \label{fig:main}
\end{figure*}

Deep image compression comprises four modules: an encoder, a quantizer, a decoder, and an entropy model.
The encoder extracts latent representations from an image.
The quantizer quantizes the latent representations for reducing bits.
The decoder reconstructs the image from the quantized representations to be close to the input.
The entropy model estimates the probabilities of the quantized representations for entropy coding. 
Deep image compression optimizes these modules in an end-to-end manner as a joint rate-distortion optimization problem.

Quantization is a critical component of image compression.
Quantization is mainly classified into two types: uniform and non-uniform.
In the traditional image compressions using orthogonal transformations, the rate-distortion optimized quantization is uniform quantization~\cite{SullivanTransInfoTheory96}.
In the deep image compression, uniform quantization is also a standard operation~\cite{conf/iclr/BalleLS17,conf/iclr/BalleMSHJ18,conf/cvpr/Cheng2020,conf/cvpr/HeCheck21}.
However, uniform quantization is defined differently during training
-- quantizers are always approximated for the end-to-end training to back-propagate meaningful gradients because the gradient of the quantization is zero almost everywhere.

There have been proposed several approximations of uniform quantization. 
Additive uniform noise (AUN-Q)~\cite{conf/iclr/BalleLS17} is the most standard approximation.
AUN-Q allows a continuous relaxation of the probability mass function of the quantized representation and approximates the quantization error.
Other studies proposed other quantization approximations such as rounding with a straight-through estimator (STE-Q)~\cite{conf/iclr/TheisSCH17}, universal quantization (U-Q)~\cite{conf/iccv/ChoiEL19,conf/nips/Agustsson20}, stochastic Gumbel annealing (SGA-Q)~\cite{conf/nips/Yang20}, and soft-then-hard quantization (STH-Q)~\cite{GuoSTHICML21}.
A combination of STE-Q for a decoder and AUN-Q for an entropy model, a.k.a. mixed approach~\cite{conf/icip/MinnenS20}, was proposed in \cite{conf/iclr/LeeCB19}.

There are no comprehensive comparisons between them.
Therefore, it is uncertain which approximated quantizations should be used. 
Some studies perform comparisons, but they are limited only to a part of them.
\cite{conf/iccv/ChoiEL19,conf/nips/Agustsson20} compared their proposed U-Q with only AUN-Q.
\cite{conf/iclr/LeeCB19} compared the combination of AUN-Q and STE-Q with only AUN-Q.
\cite{GuoSTHICML21} compared STH-Q with AUN-Q, STE-Q, and SGA-Q.
They did not compare with U-Q and the combination of AUN-Q and STE-Q, which are even used in state-of-the-art methods~\cite{CuiCVPR21,ZhuICLR22,ZouCVPR22}.

In this study, we perform exhaustive comparisons of these existing approximation methods.
We also compared their variants, a variant of SGA-Q that we named stochastic rounding with annealing (SRA-Q) and a variant of STE-Q called differentiable soft quantization (DS-Q)~\cite{conf/iccv/GongLJLHLYY19}.

Moreover, we compare the combinations of these approximation methods for a decoder and an entropy model.
The motivation is that the approximated quantizations for the decoder and the entropy model can differ.
This motivation comes from \cite{conf/iclr/LeeCB19} that showed that a combination of AUN-Q for a decoder and STE-Q for an entropy model achieves superior results than AUN-Q for both.

In the experiments, we evaluated our methods on three network architectures: Ball\'{e}17~\cite{conf/iclr/BalleLS17}, Ball\'{e}18\cite{conf/iclr/BalleMSHJ18}, and Cheng20~\cite{conf/cvpr/Cheng2020}.
We used two major datasets as test data: the Kodak dataset~\cite{Kodak} and the CLIC 2020 professional validation dataset~\cite{CLIC20}.
We also compared AUN-Q with a typical non-uniform quantization method~\cite{conf/cvpr/MentzerATTG18} and showed the superiority of uniform quantization.

Our main contributions are summarized as follows:
\begin{itemize}
  \item We compare seven approximations of uniform quantization. We also compared their combinations for a decoder and an entropy model. To the best of our knowledge, there have been no comprehensive comparisons of the approximated quantization for deep image compression.
  \item We find that there is no unique solution -- in other words, the best approximated quantization differs depending on the network architectures. 
  We also confirm that using the best approximation instead of the one used in the original deep compression improves the performance.
  \item We find that the combination of U-Q for an entropy model and DS-Q for a decoder is a comparatively good choice for different architectures and datasets.
\end{itemize}

The paper is an extended version of our conference paper~\cite{OurPaper}, which was extended in multiple aspects.
We include additional quantization approximation methods such as SGA-Q~\cite{conf/nips/Yang20}, STH-Q~\cite{GuoSTHICML21}, and DS-Q~\cite{conf/iccv/GongLJLHLYY19} in comparison,
evaluate approximated quantization in various bitrates with more datasets on more network architectures,
and describe more insights on the experimental results.

The remainder of this paper is organized as follows.
In Section~\ref{sec:related}, we give related works on deep image compression.
In Section~\ref{sec:method}, we describe the overview of our comparison.
In Section~\ref{sec:experiments}, we evaluate the combinations of approximated quantization. 
In Section~\ref{sec:conclusion}, we conclude this paper.

\section{Related Works}\label{sec:related}
\subsection{Deep Image Compression}
\begin{table*}[t]
  \centering
  \caption{List of approximated quantization. Forward denotes operation in the forward pass. Gradient denotes the gradient of approximated quantization.}
  \label{tbl:overview}
  \begin{tabular}{lcccc}
    \toprule
    Name & Approximation & Forward & Gradient & Hyper-Parameters \\
    \midrule
    AUN-Q~\cite{conf/iclr/BalleLS17} & noise & additive uniform noise (Eq.~\ref{eq:aun}) & 1 & \\
    STE-Q~\cite{conf/iclr/TheisSCH17} & rounding & deterministic rounding (Eq.~\ref{eq:ste}) & 1 & \\
    U-Q~\cite{conf/iccv/ChoiEL19} & noise & universal quantization (Eq.~\ref{eq:uq}) & 1 & \\
    SGA-Q~\cite{conf/nips/Yang20} & noise & stochastic rounding (Eq.~\ref{eq:gumbel}) & grad. of Eq.~\ref{eq:gumbel} & $c, t_0$\\
    STH-Q~\cite{GuoSTHICML21} & noise / rounding & additive uniform noise (Eq.~\ref{eq:aun}) / deterministic rounding (Eq.~\ref{eq:ste}) & 1 / 0 & $t_0$\\
    DS-Q~\cite{conf/iccv/GongLJLHLYY19} & rounding & deterministic rounding (Eq.~\ref{eq:dsq}) & Eq.~\ref{eq:dsq} & $k$\\
    SRA-Q  & rounding & stochastic rounding (Eq.~\ref{eq:sga}) & 1 & $c, t_0$\\
    \bottomrule
  \end{tabular}
\end{table*}
In the early stages, some studies addressed deep image compression that optimizes only in terms of distortion~\cite{journals/corr/TodericiOHVMBCS15,conf/cvpr/TodericiVJHMSC17,conf/cvpr/JohnstonVMCSCHS18}.
They used recurrent neural networks as the encoder and decoder and changed the bitrates by iterations during testing.
Recent studies have formulated deep image compression as a joint rate-distortion problem.
They used four modules: an entropy model, an encoder, a decoder, and a quantizer.
We present some highlights of relevant previous studies on these modules.
The review paper~\cite{journals/tcsv/MaZJZWW20} describes these more exhaustively.

The entropy model estimates the probability distribution of quantization output.
In the early stage, the probability distribution was estimated per pixel by employing a factorized-prior model~\cite{conf/iclr/BalleLS17} or Gaussian scale mixtures~\cite{conf/iclr/TheisSCH17}.
Ball\'{e} \etal~\cite{conf/iclr/BalleMSHJ18} presented a hyper-prior model that parameterized the probability distribution as a zero-mean Gaussian distribution to capture the spatial redundancy of the quantization output.
The parameters of the probability distribution were estimated by another auto-encoder.
Minnen \etal~\cite{conf/nips/MinnenBT18} extended \cite{conf/iclr/BalleMSHJ18} by proposing an autoregressive context model that predicted the parameters of the probability distribution from the quantized representations.
Cheng \etal~\cite{conf/cvpr/Cheng2020} extended \cite{conf/nips/MinnenBT18} by introducing a Gaussian mixture model instead of a single Gaussian model.
He \etal~\cite{conf/cvpr/HeCheck21} used a checkerboard context model for parallel computations.
Guo \etal~\cite{guo2021causal} proposed a causal context model that captures global-scope spatial relationships and cross-channel relationships.

Many studies used architectures proposed in image super-resolution to improve the encoder and decoder.
Li \etal~\cite{conf/cvpr/LiZGZ018} removed the batch normalization layer~\cite{BatchNorm} from the residual blocks~\cite{conf/cvpr/HeZRS16} in their experiments.
Attention-based architectures such as channel attention~\cite{ChannelAttn} and non-local attention~\cite{NonLocalAttn} were also used in \cite{conf/ijcai/ZhongAA20} and \cite{conf/cvpr/Cheng2020,arxiv/Liu-1904-09757,AttentionTIP21}, respectively.
Some studies proposed task-specific architectures for deep image compression.
Lin \etal~\cite{conf/cvpr/Lin2020} proposed a spatial recurrent neural network to remove redundant information between adjacent blocks.
Wang \etal~\cite{wang2020ensemble} addressed an ensemble of encoders and entropy models in deep image compression with block-wise coding. 

These methods use quantization in their pipeline regardless of the architecture of the entropy model, encoder, and decoder.
In this study, we focus on a quantizer and detail related works on quantization.

\subsection{Quantizer in Deep Image Compression}
Quantization is classified into uniform quantization and non-uniform quantization.
Similar to traditional compression methods such as JPEG~\cite{JPEG}, uniform quantization is generally used in deep image compression.

AUN-Q~\cite{conf/iclr/BalleLS17} has been the most widely used approximation~\cite{conf/cvpr/Cheng2020,conf/iclr/TheisSCH17,conf/iclr/BalleMSHJ18,conf/nips/MinnenBT18} since it was proposed in \cite{conf/iclr/BalleLS17}.
Choi \etal~\cite{conf/iccv/ChoiEL19} introduced universal quantization~\cite{UniversalQuantizationZiv85,UniversalQuantization} for approximated quantization.
They proposed U-Q as another alternative for AUN-Q and achieved better performance than AUN-Q.
Theis \etal~\cite{conf/iclr/TheisSCH17} proposed STE-Q, which extended binarization with STE~\cite{conf/nips/HubaraCSEB16} for deep image compression.
Yang \etal~\cite{conf/nips/Yang20} proposed SGA-Q to refine the quantized latent representations by iterative inference.
Guo \etal~\cite{GuoSTHICML21} presented STH-Q.
They firstly trained with AUN-Q.
Then, they trained only a decoder and an entropy model by performing quantization without approximation to reduce the mismatch between training and testing.

In general, the approximated quantization is the same for the decoder and the entropy model; however, they can be different.
In this study, we evaluate the combinations of different approximated quantization.
\cite{conf/iclr/LeeCB19} is so far the only work that used different approximated quantizers.
They used STE-Q for the decoder and AUN-Q for the entropy model.
They achieved better performance than only using AUN-Q.
This combination resulted in an improvement of the compression performance~\cite{conf/iclr/LeeCB19,conf/icip/MinnenS20,MentzerHific}.
We consider that the effective approximation method differs between the decoder and the entropy model.
Therefore, we comprehensively compare the combinations of several approximation methods for a decoder and an entropy model.

Pan \etal~\cite{conf/cvprw/Pan21} studied the mechanism of the approximated quantization.
They identified three gaps in the approximated quantization: discrete gap, entropy estimation gap, and local smoothness gap.
They analyzed these gaps and addressed them by proposing soft-STE.
In contrast, our study focuses on an empirical comparison of existing methods and their combinations.

Non-uniform quantization uses a non-uniform quantization interval instead of a uniform quantization interval.
Some studies based on clustering techniques~\cite{conf/nips/AgustssonMTCTBG17,conf/cvpr/MentzerATTG18,conf/ijcai/ZhongAA20} learned the quantization interval.
Cai \etal~\cite{conf/icip/CaiZ18} alternatively optimized the quantization interval and the encoder-decoder.
Although several studies have used non-uniform quantization, non-uniform quantization is not standard in current deep image compression studies (\eg, a state-of-the-art compression method~\cite{ZouCVPR22} uses uniform quantization).
Therefore, we mainly focus our comparison to uniform quantization.

A few works investigated other quantization.
Li \etal~\cite{conf/dcc/LiAL020} proposed to incorporate trellis coded quantization.
L\"ohdefink \etal~\cite{Lohdefink_2022_CVPR} proposed a one-hot max quantization.
Although these methods showed better performance at low rates such as 0.1 bits per pixel (BPP), they perform worse at rates higher than 0.45 BPP.

\section{Method}\label{sec:method}
We compare approximated quantization and their combinations comprehensively.
We first explain the outline of deep image compression in our comparison.
Thereafter, we explain existing approximated quantization and the variants that we compare.
Then, we explain their combination for a decoder and an entropy model.
Finally, we explain the implementation of quantization for compression models using a hyper-prior model~\cite{conf/iclr/BalleMSHJ18} as the entropy model.

\subsection{Outline of Our Comparison}
Image compression aims to compress an image into a small number of bits and reconstruct the image from them.
Deep image compression achieves this goal using four modules: a decoder, an encoder, an entropy model, and a quantizer.
In our comparison, we use two quantizers for an entropy model and a decoder instead of a single quantizer.
We show the outline in Fig.~\ref{fig:main}.
Given quantized latent representations extracted by an encoder and two quantizers from an image, a decoder reconstructs the image, and an entropy model estimates the probabilities for entropy coding.
These modules are learned in an end-to-end manner using a joint rate-distortion optimization framework.

\begin{figure}[t]
  \centering
  \includegraphics[width=\hsize]{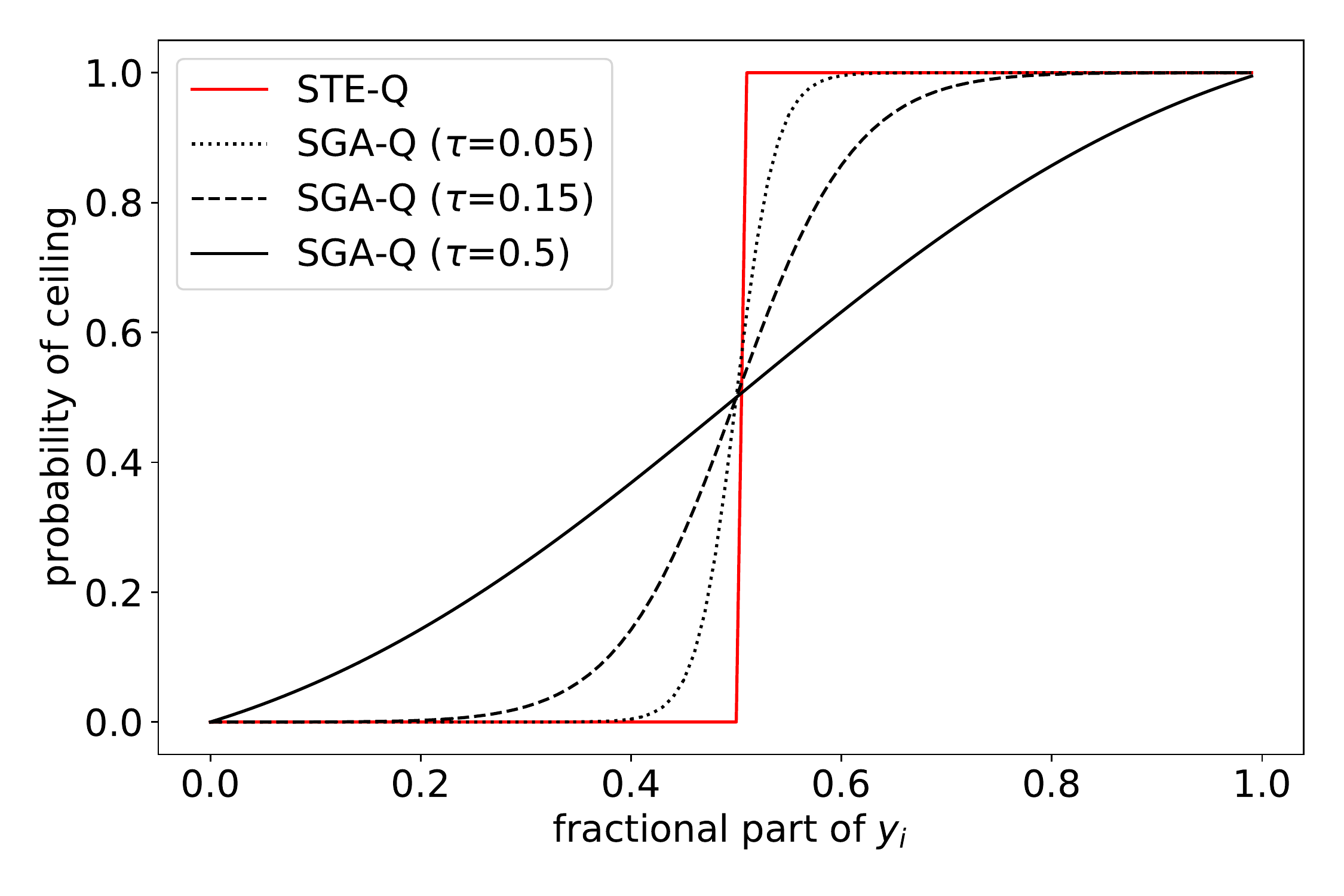}
  \caption{Visualization of SGA-Q compared to STE-Q. $\tau$ gradually reduces in training and SGA-Q becomes closer to STE-Q.}
  \label{fig:sgaq}
\end{figure}

We explain the outline of deep image compression in our comparison with mathematical formulations.
Let $\x \in \mathbb{R}^{N}$ be a vector of the original image, $f: \mathbb{R}^N \to \mathbb{R}^M$ be an encoder, $q^{\{\mathrm{ent,dec}\}}: \mathbb{R}^M \to \mathbb{R}^M$ be two quantizers for an entropy model and a decoder, and $g: \mathbb{R}^M \to \mathbb{R}^N$ be a decoder.
$N \in \mathbb{Z}$ is $HWC$, where $H, W, C \in \mathbb{Z}$ are the height, width, and number of channels of the original image, respectively.
$M \in \mathbb{Z}$ is $H'W'C'$, where $H', W', C' \in \mathbb{Z}$ are those of the latent representations.

Deep image compression aims to minimize the distortion and the rate jointly.
Let $D: \mathbb{R}^N \times \mathbb{R}^N \to \mathbb{R}$ be the distortion, $R: \mathbb{R}^M \to \mathbb{R}$ be the rate, and $\lambda \in \mathbb{R}$ be a hyper-parameter to balance the output of $D$ and $R$.
The loss function is written as
\begin{equation}
  \label{eq:rd}
  \mathcal{L} = R (q^{\mathrm{ent}}(f(\x))) + \lambda D (\x, g(q^{\mathrm{dec}}(f(\x)))).
\end{equation}
If $q^{\mathrm{ent}} = q^{\mathrm{dec}}$, we use the same approximated quantization for an entropy model and a decoder in our implementation.
It is equal to using only a single quantizer.

During testing, $q^{\{\mathrm{ent,dec}\}}$ becomes a quantizer that is a round function.
Therefore, the quantization during testing does not depend on the approximation method during the training time.

\subsection{Details of Existing Quantization Methods}
Quantization should be differentiable to back-propagate meaningful gradients for end-to-end learning.
We explain five existing approximated quantizations and their two variants that we compare in our experiments.
We list them in Table~\ref{tbl:overview}.
The five existing methods are AUN-Q, STE-Q, U-Q, SGA-Q, and STH-Q.
The two variants are DS-Q and SRA-Q.
Let $\y = f(\x) \in \mathbb{R}^M$ and $\tilde{\y}^{\{\mathrm{ent,dec}\}} = q^{\{\mathrm{ent,dec}\}}(\y) \in \mathbb{R}^M$ to explain each method with mathematical formulations.

\textbf{AUN-Q}~\cite{conf/iclr/BalleLS17} approximates quantization by adding uniform noise. 
Let $i \in \{1, \dots, M\}$ and $u_i \in U[-\frac{1}{2}, \frac{1}{2}]$, where $U$ is a uniform distribution.
AUN-Q is written as follows:
\begin{equation}\label{eq:aun}
  \tilde{y}_i^{\{\mathrm{ent,dec}\}} = y_i + u_i.
\end{equation}
The gradient is one.
AUN-Q approximates the quantization operation well in the probability distribution.
However, there is an apparent gap between training and testing; noise is not added during testing.
This is called the discrete gap in \cite{conf/cvprw/Pan21}.

\begin{figure}[t]
  \centering
  \includegraphics[width=\hsize]{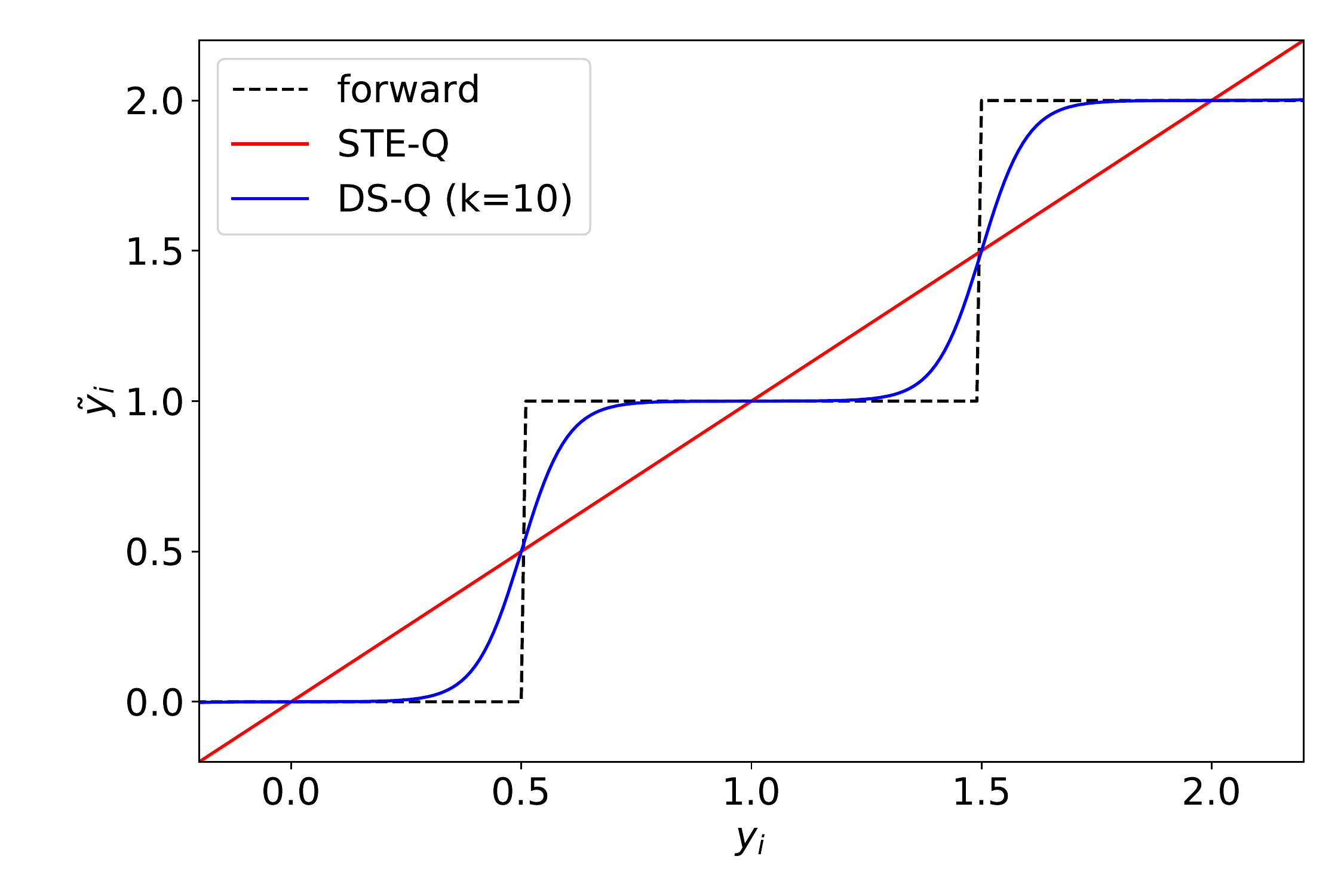}
  \caption{Visualization of DS-Q compared to STE-Q.}
  \label{fig:dsq}
\end{figure}

\textbf{STE-Q}~\cite{conf/iclr/TheisSCH17} approximates the gradient of quantization using STE~\cite{conf/nips/HubaraCSEB16}.
The forward pass is quantization without approximation and is written as
\begin{equation}\label{eq:ste}
  \tilde{y}_i^{\{\mathrm{ent,dec}\}} = \lfloor y_i \rceil,
\end{equation}
where $\lfloor \cdot \rceil$ is the round function.
For back-propagation, STE assumes the forward pass is the identity function; therefore, the gradient is one.
The advantage of STE-Q is that the operation during training is equivalent to that during testing.
STE-Q does not learn the probability distribution of quantized representations.
This is called the entropy estimation gap in \cite{conf/cvprw/Pan21}.

\begin{table*}[t]
  \centering
  \caption{Network architectures in our experiments.}
  \label{tbl:architectures}
  \begin{tabular}{lccc}
    \toprule
    Name & Encoder/Decoder & Entropy Model & Quantizer\\
    \midrule
    Ball\'{e}17~\cite{conf/iclr/BalleLS17} & three convolutional layers & factorized-prior & AUN-Q\\
    Ball\'{e}18~\cite{conf/iclr/BalleMSHJ18} & four convolutional layers & hyper-prior & AUN-Q\\
    Cheng20~\cite{conf/cvpr/Cheng2020} & residual blocks and simplified attention blocks & hyper-prior and autoregressive context & AUN-Q\\
    \bottomrule
  \end{tabular}
\end{table*}

\textbf{U-Q} in \cite{conf/iccv/ChoiEL19} applies universal quantization~\cite{UniversalQuantizationZiv85,UniversalQuantization} to deep image compression.
Given a common uniform noise $u \in U[-\frac{1}{2}, \frac{1}{2}]$, U-Q is written as follows:
\begin{equation}\label{eq:uq}
  \tilde{y}_i^{\{\mathrm{ent,dec}\}} = \lfloor y_i + u \rceil - u.
\end{equation}
The probability density functions of U-Q and AUN-Q are the same.
The gradient of U-Q is approximated by STE, and the gradient is one.
U-Q achieves better performance than AUN-Q.

\textbf{SGA-Q}~\cite{conf/nips/Yang20} perform rounding stochastically instead of deterministically.
SGA-Q becomes more deterministic in the latter part of training.
We visualize SGA-Q in Fig.~\ref{fig:sgaq}.
Let $\tau = \min (0.5, 0.5 \exp(-c(t - t_0))) \in \mathbb{R}$ where $t \in \mathbb{Z}$ is the iteration and $c \in \mathbb{R}$ and $t_0 \in \mathbb{Z}$ are hyper-parameters to adjust $\tau$.
SGA-Q can be written as
\begin{equation}\label{eq:sga}
  \tilde{y}_i^{\{\mathrm{ent,dec}\}} = \lfloor y_i \rfloor + \delta, \\
\end{equation}
where $\delta \in \{0, 1\}$.
$P(\delta = 0) = 1 - p_{\tau} \propto \mathrm{exp} (-\mathrm{arctanh} (y_i - \lfloor y_i \rceil) / \tau)$ and $P(\delta = 1) = p_{\tau} \propto \mathrm{exp} (-\mathrm{arctanh} (\lfloor y_i \rceil - y_i) / \tau)$.
As the number of iterations increases, $\tau$ reduces gradually and $P(\delta = 0)$ and $P(\delta = 1)$ become close to zero or one.
SGA-Q behaves stably around the boundary of rounding, in contrast to STE-Q.
We approximate the gradient using the Gumbel-softmax trick~\cite{Gumbel/Jang,Gumbel/Maddison}.
Yang \etal~\cite{conf/nips/Yang20} used this technique only for iterative inference, but we used it for training.
By using the Gumbel-softmax trick, the forward pass is rewritten as follows:

\begin{align}\label{eq:gumbel}
  \tilde{y}_i^{\{\mathrm{ent,dec}\}} &= \lfloor y_i \rfloor \nonumber \\
  &+ \frac{h_{\tau}(\log p_{\tau} + g_0)}{h_{\tau}(\log p_{\tau} + g_0) + h_{\tau}(\log(1 - p_{\tau}) + g_1)},
\end{align}
where $g_0$ and $g_1$ are sampled from the Gumbel(0, 1) distribution and $h_{\tau} (x) = \exp(x / \tau)$.
The gradient is calculated using this forward pass.
Because SGA-Q adds noise drawn from a Gumbel distribution, it is considered as a noise-based method.

In \textbf{STH-Q}~\cite{GuoSTHICML21}, a compression model is first trained using AUN-Q.
Thereafter, it disables the approximation of the quantization at the $t_0$-th iteration to train a decoder and an entropy model, where $t_0$ is a hyper-parameter.
Therefore, the gradient is one before the $t_0$-th iteration and zero thereafter.
The advantage of this approach is that there is no quantization gap between the training time after the change and the testing time.
However, the encoder can be suboptimal because it is not trained jointly with the decoder and the entropy model in the latter part of the training.

\textbf{DS-Q}~\cite{conf/iccv/GongLJLHLYY19} is a variant of STE-Q.
It calculates the gradient assuming the forward pass uses the tanh function.
While STE-Q calculates the gradient assuming that the forward pass is the identity function, this causes a gradient mismatch.
The gradient mismatch leads to a suboptimal solution.
DS-Q reduces this gradient mismatch by approximating it via a continuous function closer to the round function.
We visualize DS-Q in Fig.~\ref{fig:dsq}.
The equation is written as follows:
\begin{align}\label{eq:dsq}
  \frac{\partial}{\partial y_i} \tilde{y}_i^{\{\mathrm{ent,dec}\}} = \frac{\partial}{\partial y_i} \left[\lfloor y_i \rfloor + \frac{1}{2} + \frac{1}{2} \frac{\mathrm{tanh}(kd_i)}{\mathrm{tanh}(0.5k)} \right],\\
  \mathrm{where}~d_i = y_i - \lfloor y_i \rfloor - 0.5.
\end{align}
The approximated quantization becomes close to the round function as $k$ increases.
It becomes close to the identity function if $k$ is small.
We treat $k$ as a hyper-parameter following \cite{conf/nips/Agustsson20}.
We do not increase $k$ gradually as \cite{conf/nips/Agustsson20} -- we empirically compare increasing and fixing $k$ and adopt the better.
In \cite{conf/iccv/GongLJLHLYY19}, $k$ was treated as a learnable parameter, but this is not successful in our experiments.

\textbf{SRA-Q} is a variant of SGA-Q.
In the forward pass, SRA-Q performs stochastic rounding as shown in Eq.~\ref{eq:sga}.
SRA-Q calculates the gradient following the stochastic rounding in \cite{conf/iclr/TheisSCH17}, where the gradient becomes one after calculating it by its expectation.

\subsection{Combination of Approximation Methods for Two Quantizers}
We apply these seven approximation methods to two quantizers separately.
Specifically, we evaluate seven approximated quantization methods and pairs of six approximation methods for an entropy model and a decoder.
We exclude STH-Q in making pairs because it is not separable for an entropy model and a decoder.
The total number of approximated quantization that we evaluate is $7 + {}_6 P_2 = 37$ in total.

\subsection{Quantization for a Hyper-prior Model}
A hyper-prior model~\cite{conf/iclr/BalleMSHJ18} applies an auto-encoder to latent representations to capture the redundancy of the latent representations.
The extracted features by the auto-encoder are called hyper latents.
They are quantized to compress by entropy coding like the latent representations.
We used the same quantization method to the hyper latents following Lee \etal~\cite{conf/iclr/LeeCB19}.
The amount of this overhead is negligible compared to the latent representation.

\section{Experiments}
\label{sec:experiments}

\begin{table}[t]
  \centering
  \caption{Comprehensive comparison of approximated quantization using the network architecture of Ball\'{e}17~\cite{conf/iclr/BalleLS17}. The best one is shown in bold, and the second-best one is shown with an underline. The original Ball\'{e}17 uses AUN-Q, marked with *.}
  \label{tbl:comparison_balle17}
  \begin{tabular}{cc|ccc}
    \toprule
    Entropy & \multirow{2}{*}{Decoder} & \multicolumn{3}{c}{BD rate (\%) $\downarrow$}\\
    Model & & Kodak & CLIC & Average\\
    \midrule
    \multicolumn{4}{c}{Single Quantizer}\\
    \midrule
    \multicolumn{2}{c|}{AUN-Q*} &     0.00 &     0.00 & 0.00\\
    \multicolumn{2}{c|}{STE-Q} &    9.10 & 9.18 & 9.14\\
    \multicolumn{2}{c|}{U-Q} &    -6.59 & -8.08 & -7.34\\
    \multicolumn{2}{c|}{SGA-Q} &    -5.32 & -5.85 & -5.58\\
    \multicolumn{2}{c|}{DS-Q} &     13.13 &     13.25 & 13.19\\
    \multicolumn{2}{c|}{SRA-Q} &    -1.71 & -1.78 & -1.74\\
    \multicolumn{2}{c|}{STH-Q}  &    -7.90 & -8.77 & -8.33\\
    \midrule
    \multicolumn{4}{c}{Combination of Quantizers}\\
    \midrule
    AUN-Q & STE-Q &   \underline{-8.84} &    -9.41 & -9.12\\
    AUN-Q & U-Q &    -7.47 &    -8.78 & -8.13\\
    AUN-Q & SGA-Q &    -6.18    &    -8.44 & -7.31\\
    AUN-Q & DS-Q &    -6.87    &    -7.03 & -6.95\\
    AUN-Q & SRA-Q &    -8.22    &    -9.67 & -8.95\\
    STE-Q & AUN-Q &    14.85 &    14.02 & 14.44\\
    STE-Q & U-Q &    10.11 & 8.83 & 9.47\\
    STE-Q & SGA-Q &    15.26 &    13.43 & 14.35\\
    STE-Q & DS-Q &    12.15 &    14.21 & 13.18\\
    STE-Q & SRA-Q &     10.63 &     12.40 & 11.51\\
    U-Q & AUN-Q &    -1.84 &    -2.13 & -1.99\\
    U-Q & STE-Q &    -8.30&    -8.91 & -8.61\\
    U-Q & SGA-Q &    -5.57&    -7.31 & -6.44\\
    U-Q & DS-Q &    \textbf{-8.93} &    \underline{-9.78} & \underline{-9.35}\\
    U-Q & SRA-Q &    -8.73 &    \textbf{-10.04} & \textbf{-9.38}\\
    SGA-Q & AUN-Q & -0.68 & -0.91 & -0.79\\
    SGA-Q & STE-Q &    -7.69 & -8.36 & -8.03\\
    SGA-Q & U-Q &     0.59 &     0.60 & 0.59\\
    SGA-Q & DS-Q &    -7.57 & -8.00 & -7.79\\
    SGA-Q & SRA-Q &    -7.71 & -8.20 & -7.96\\
    DS-Q & AUN-Q &    14.61 &    13.66 & 14.14\\
    DS-Q & STE-Q &     10.95 &     10.89 & 10.92\\
    DS-Q & U-Q &     9.89 &     7.97 & 8.93\\
    DS-Q & SGA-Q &    15.44 &    13.03 & 14.23\\
    DS-Q & SRA-Q &    24.36 &    25.04 & 24.70\\
    SRA-Q & AUN-Q &     6.25 &     5.20 & 5.72\\
    SRA-Q & STE-Q &    -2.79 &    -1.90 & -2.35\\
    SRA-Q & U-Q &    -1.53&    -2.01 & -1.77\\
    SRA-Q & SGA-Q &     0.32 &    -0.22 & 0.05\\
    SRA-Q & DS-Q &    -2.59 &    -1.82 & -2.20\\
    \bottomrule
  \end{tabular}
\end{table}

\subsection{Experimental Settings}\label{subsec:exp_setup}
Regarding the training data, we used a subset of ImageNet~\cite{ImageNet}.
We used high-resolution images whose shorter edge is larger than 256 pixels.
We cropped the images at random locations to obtain patches with a resolution of $256 \times 256$ pixels.
We used two datasets for the test data: the Kodak dataset~\cite{Kodak} and the CLIC 2020 professional validation dataset (CLIC)~\cite{CLIC20}.
The Kodak dataset comprises 24 images.
The resolution of the images is $512 \times 768$ or $768 \times 512$ pixels.
The CLIC dataset comprises 41 images.
The resolution of the many images is approximately $2,048\times1,360$ pixels.

We made evaluations using three network architectures: Ball\'{e}17~\cite{conf/iclr/BalleLS17}, Ball\'{e}18~\cite{conf/iclr/BalleMSHJ18}, and Cheng20~\cite{conf/cvpr/Cheng2020}.
We summarized the network architectures in Table~\ref{tbl:architectures}.
The latter increases the complexity of the architecture.
Specifically, Cheng20 uses six and seven residual blocks~\cite{conf/cvpr/HeZRS16} for the encoder and decoder, respectively.
They used simplified non-local attention modules in these layers.
The entropy model is based on Gaussian mixtures, whose parameters are predicted by a hyper-prior and an autoregressive context model.
The quantizer is approximated using AUN-Q.
Some convolutional layers in the encoder and decoder of these three networks are followed by generalized divisible normalization~\cite{Balle/iclr/GDN16}.

We used Adam optimizer~\cite{kingma2015adam}.
The number of iterations was 1,000,000 and the size of the mini-batch was eight.
We set the initial learning rate to $10^{-4}$.
For Cheng20, we set the learning rate to $10^{-5}$ for the last 80,000 iterations.
The distortion function was a mean squared error.
To prepare compression models with various bitrates, we set $\lambda$ to $\{0.001, 0.003, 0.01, 0.03\}$ for Ball\'{e}17 and Ball\'{e}18, and \{0.0016, 0.003, 0.0075, 0.15, 0.30, 0.45\} for Cheng20.

We used Bj{\o}ntegaard Delta bitrate (BD rate)~\cite{BDrate} as a metric.
This metric indicates the bitrate savings compared to a baseline method in the same distortion.
We evaluated the distortion by the peak signal-to-noise ratio (PSNR) and the bitrate by BPP to compute the BD rate.
In our experiments, we treated compression models using AUN-Q as the baseline method for each network architecture.

We tuned hyper-parameters for SGA-Q, SRA-Q, DS-Q, and STH-Q on Ball\'{e}17~\cite{conf/iclr/BalleLS17} on the Kodak dataset.
We set $c = 0.0003, t_0 = 960,000$ for SGA-Q, $c = 0.0003, t_0 = 990,000$ for SRA-Q, and $t_0 = 960,000$ for STH-Q.
We fixed $k$ to $0.1$ for DS-Q because increasing $k$ gradually does not improve the performance.
For the reproducibility of our experiments, we provide our code at \url{https://github.com/kktsubota/uniform-quantizers}.

\begin{table}[t]
  \centering
  \caption{Comparison of the approximated quantization using the network architecture of Ball\'{e}18~\cite{conf/iclr/BalleMSHJ18}. The best one is shown in bold, and the second-best one is shown with an underline. The original Ball\'{e}18 uses AUN-Q, marked with *.}
  \label{tbl:comparison_balle18}
  \begin{tabular}{cc|ccc}
    \toprule
    Entropy & \multirow{2}{*}{Decoder} & \multicolumn{3}{c}{BD rate (\%) $\downarrow$}\\
    Model & & Kodak & CLIC & Average\\
    \midrule
    \multicolumn{4}{c}{Single Quantizer}\\
    \midrule
    \multicolumn{2}{c|}{AUN-Q*}  &      0.00 &     0.00 & 0.00\\
    \multicolumn{2}{c|}{U-Q} &   -5.13 &    -6.14 & -5.64\\
    \multicolumn{2}{c|}{SGA-Q}  &     3.20 &     1.42 & 2.31\\
    \multicolumn{2}{c|}{STH-Q}  &    -4.74 &    -6.06 & -5.40\\
    \midrule
    \multicolumn{4}{c}{Combination of Quantizers}\\
    \midrule
    AUN-Q & STE-Q &    \underline{-6.09} &    \underline{-7.17} & \underline{-6.63}\\
    AUN-Q & U-Q &    -5.26 &    -5.74 & -5.50\\
    AUN-Q & SGA-Q &     -0.97 &    -2.90 & -1.93\\
    AUN-Q & DS-Q &    \textbf{-6.10} &    \textbf{-7.77} & \textbf{-6.94}\\
    AUN-Q & SRA-Q &    -3.27 &    -4.03 & -3.65\\
    U-Q & STE-Q &   -5.37 &    -7.28 & -6.33\\
    U-Q & SGA-Q &    0.60 &    -1.05 & -0.23\\
    U-Q & DS-Q &   -5.41 & -6.09 & -5.75\\
    U-Q & SRA-Q &    -2.00 &    -1.31 & -1.66\\
    SGA-Q & STE-Q &   0.30 &    -1.61 & -0.66\\
    SGA-Q & DS-Q &     -1.53 &    -4.02 & -2.77\\
    SGA-Q & SRA-Q &    0.85 &     1.21 & 1.03\\
    \bottomrule
  \end{tabular}
\end{table}

\begin{table}[t]
  \centering
  \caption{Comparison of the approximated quantization using the network architecture of Cheng20~\cite{conf/cvpr/Cheng2020}. The best one is shown in bold, and the second-best one is shown with an underline. The original Cheng20 uses AUN-Q, marked with *.}
  \label{tbl:comparison_cheng20}
  \begin{tabular}{cc|ccc}
    \toprule
    Entropy & \multirow{2}{*}{Decoder} & \multicolumn{3}{c}{BD rate (\%) $\downarrow$}\\
    Model & & Kodak & CLIC & Average\\
    \midrule
    \multicolumn{2}{c|}{AUN-Q*} & 0.00 & 0.00 & 0.00\\
    \multicolumn{2}{c|}{U-Q} & \underline{-2.55} & -2.51 & -2.53\\
    \multicolumn{2}{c|}{STH-Q} & -2.28 & \underline{-3.14} & \underline{-2.71}\\
    AUN-Q & STE-Q & -1.03 & -0.99 & -1.01\\
    AUN-Q & U-Q & \textbf{-2.95} & \textbf{-3.34} & \textbf{-3.15}\\
    AUN-Q & DS-Q & 0.12 & 0.67 & 0.40\\
    U-Q & STE-Q & -0.78 & -0.12 & -0.45\\
    U-Q & DS-Q & -1.96 & -1.99 & -1.98\\
    \bottomrule
  \end{tabular}
\end{table}

\begin{figure*}[t]
  \begin{minipage}{0.32\hsize}
      \centering
      \includegraphics[width=\hsize]{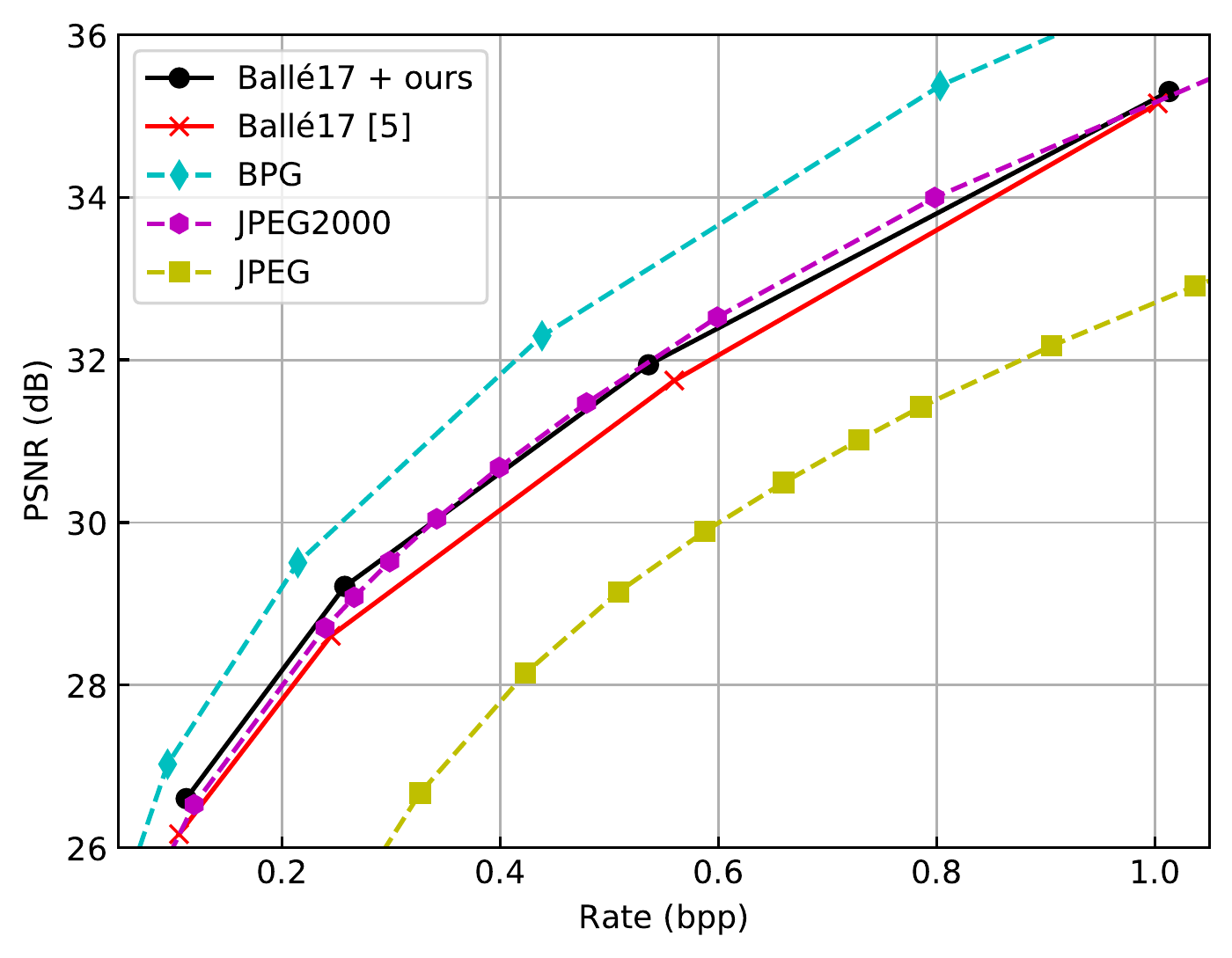}
      \subcaption{Ball\'{e}17}
  \end{minipage}
  \begin{minipage}{0.32\hsize}
      \centering
      \includegraphics[width=\hsize]{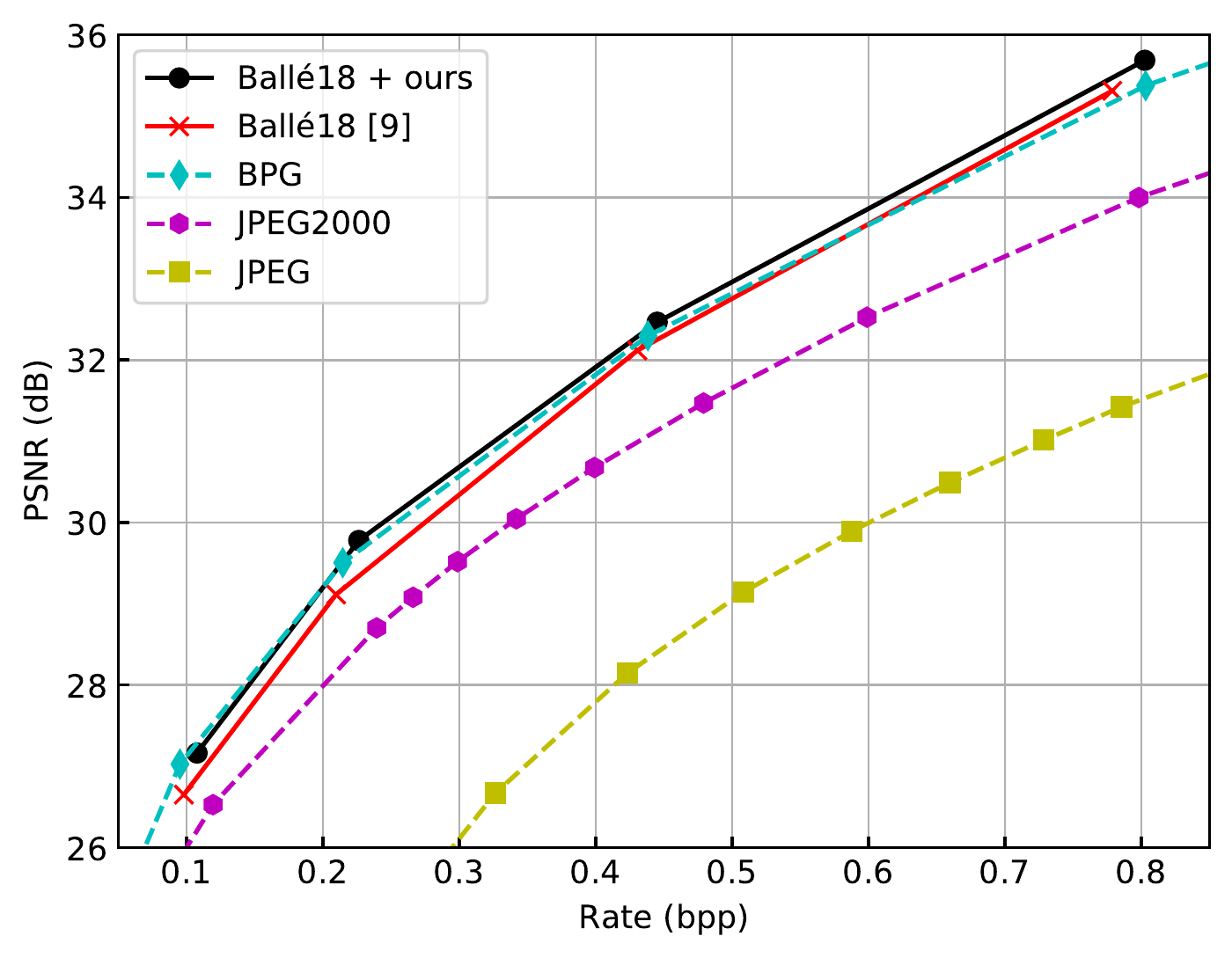}
      \subcaption{Ball\'{e}18}
  \end{minipage}
  \begin{minipage}{0.32\hsize}
      \centering
      \includegraphics[width=\hsize]{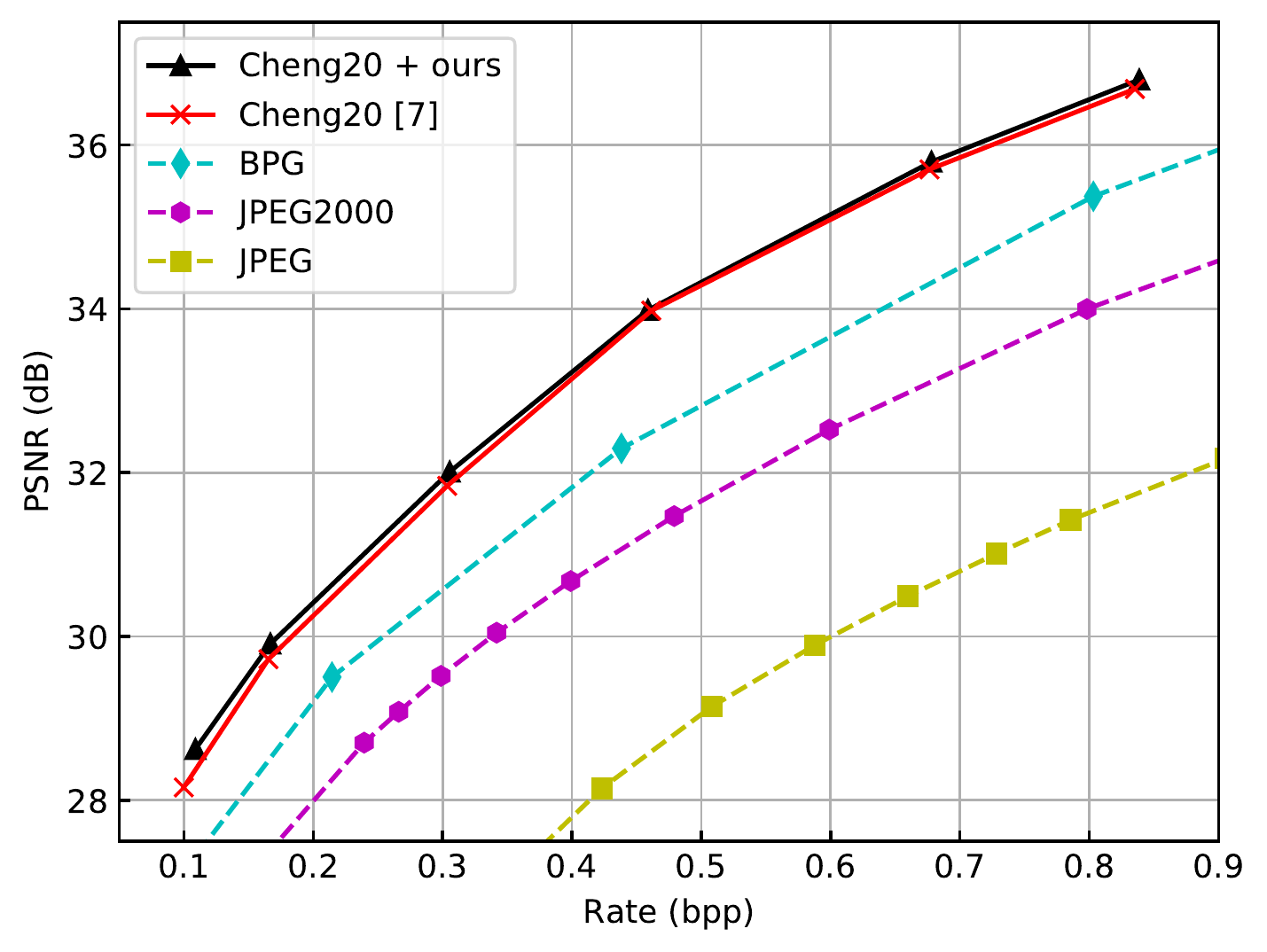}
      \subcaption{Cheng20}
  \end{minipage}
  \caption{Rate-distortion curve on the Kodak dataset. Ball\'{e}17 + ours, Ball\'{e}18 + ours, and Cheng20 + ours denote Ball\'{e}17~\cite{conf/iclr/BalleLS17}, Ball\'{e}18~\cite{conf/iclr/BalleMSHJ18}, and Cheng20~\cite{conf/cvpr/Cheng2020} using the best approximation, respectively.}
  \label{fig:comparison}
\end{figure*}

\begin{table}[t]
  \centering
  \caption{Average BD rate of Ball\'{e}17, Ball\'{e}18, and Cheng20. The best one is shown in bold, and the second-best one is shown with an underline.}
  \label{tbl:average}
  \begin{tabular}{cc|ccc}
    \toprule
    Entropy & \multirow{2}{*}{Decoder} & \multicolumn{3}{c}{BD rate (\%) $\downarrow$}\\
    Model & & Kodak & CLIC & Average\\
    \midrule
    \multicolumn{2}{c|}{AUN-Q} & 0.00 & 0.00 & 0.00\\
    \multicolumn{2}{c|}{U-Q} & -4.76 & -5.58 & -5.17\\
    \multicolumn{2}{c|}{STH-Q} & -4.97 & \textbf{-5.99} & -5.48\\
    AUN-Q & STE-Q & \underline{-5.32} & -5.86 & \underline{-5.59}\\
    AUN-Q & U-Q & -5.23 & -5.95 & \underline{-5.59}\\
    AUN-Q & DS-Q & -4.28 & -4.71 & -4.50\\
    U-Q & STE-Q & -4.82 & -5.44 & -5.13\\
    U-Q & DS-Q & \textbf{-5.43} & \underline{-5.95} & \textbf{-5.69}\\
    \bottomrule
  \end{tabular}
\end{table}

We first performed experiments using Ball\'{e}17 for the exhaustive cases, and then using Ball\'{e}18 and Cheng20 for limited cases to reduce experimental costs.
For Ball\'{e}18, we conducted experiments using approximations whose BD rate in the Kodak dataset is less than -4\% in Ball\'{e}17.
For Cheng20, we conducted experiments using approximations whose BD rate in the Kodak dataset is less than -4\% in Ball\'{e}18.
In experiments, we train these models two times and reported the average score.

\subsection{Evaluation of Approximated Quantization}
We report the results of Ball\'{e}17~\cite{conf/iclr/BalleLS17}, Ball\'{e}18~\cite{conf/iclr/BalleMSHJ18}, and Cheng20~\cite{conf/cvpr/Cheng2020} in Table~\ref{tbl:comparison_balle17}, \ref{tbl:comparison_balle18}, and \ref{tbl:comparison_cheng20}, respectively.
We find that using the best approximation for each network architecture outperforms its original approximation methods.
The improvement in BD rate is -9.38\%, -6.94\%, and -3.15\%, respectively.
We observe that the improvement becomes less when the network architecture is more powerful.
The best approximation for each network architecture is different from the approximation methods that have been proposed in previous studies.

The best approximation differs depending on the network architectures.
The combination of U-Q for an entropy model and SRA-Q for a decoder is best for Ball\'{e}17, whereas the combination is not best for Ball\'{e}18 and Cheng20.
The best approximation for Ball\'{e}18 is the combination of AUN-Q and DS-Q, and the best combination for Cheng20 is the combination of AUN-Q and U-Q.
The results indicate that we might need to explore approximation methods per network architecture for the best approximation.

We calculated the average BD rate over the three network architectures and reported it in Table~\ref{tbl:average}.
The results show that the combination of U-Q for an entropy model and DS-Q for a decoder is comparatively good among them.

We finally discuss the tendency of good approximation methods.
In Table~\ref{tbl:comparison_balle17}, we can observe that using noise-based methods such as AUN-Q, U-Q, and SGA-Q for an entropy model and rounding-based methods such as STE-Q, DS-Q, and SRA-Q for a decoder leads to better performance.
When using a single quantizer, noise-based methods such as AUN-Q, U-Q, and SGA-Q are better than deterministic rounding-based methods such as STE-Q and DS-Q.
It indicates that the entropy estimation gap~\cite{conf/cvprw/Pan21} is more important than the discrete gap~\cite{conf/cvprw/Pan21}.

\subsection{Performance of the Best Approximated Quantization}
\begin{figure*}[t]
  \begin{minipage}{0.32\hsize}
    \centering
    \begin{minipage}{0.73\hsize}
      \centering
      \includegraphics[width=\hsize]{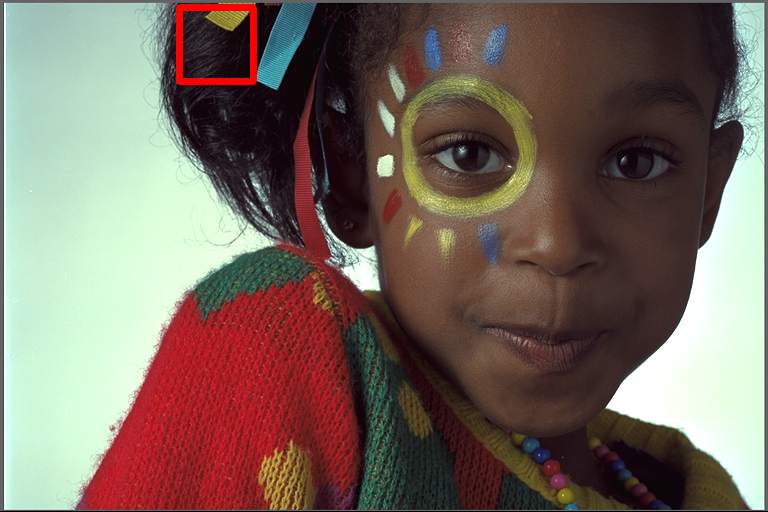}
    \end{minipage}
    \begin{minipage}{0.25\hsize}
      \begin{minipage}{0.9\hsize}
        \centering
        \setlength{\fboxrule}{1pt}
        \setlength{\fboxsep}{0pt}
        \fcolorbox{red}{white}{\includegraphics[width=\hsize]{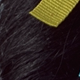}}
      \end{minipage}\\
      \begin{minipage}{0.9\hsize}
      \end{minipage}\\
      \begin{minipage}{0.9\hsize}
        \centering
        \setlength{\fboxrule}{1pt}
        \setlength{\fboxsep}{0pt}
        \fcolorbox{red}{white}{\includegraphics[width=\hsize]{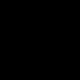}}
      \end{minipage}
    \end{minipage}
    \subcaption*{Original Image\\PSNR / BPP}
  \end{minipage}
  \begin{minipage}{0.32\hsize}
    \centering
    \begin{minipage}{0.73\hsize}
      \centering
      \includegraphics[width=\hsize]{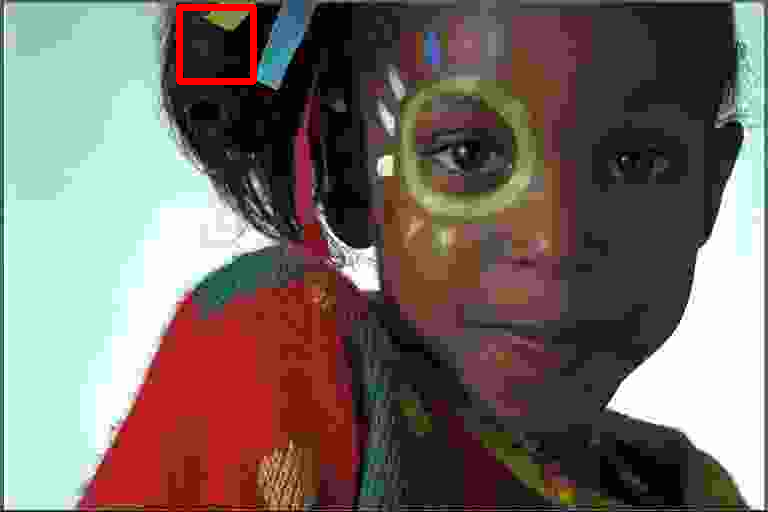}
    \end{minipage}
    \begin{minipage}{0.25\hsize}
      \begin{minipage}{0.9\hsize}
        \centering
        \setlength{\fboxrule}{1pt}
        \setlength{\fboxsep}{0pt}
        \fcolorbox{red}{white}{\includegraphics[width=\hsize]{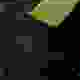}}
      \end{minipage}\\
      \begin{minipage}{0.9\hsize}
      \end{minipage}\\
      \begin{minipage}{0.9\hsize}
        \centering
        \setlength{\fboxrule}{1pt}
        \setlength{\fboxsep}{0pt}
        \fcolorbox{red}{white}{\includegraphics[width=\hsize]{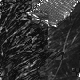}}
      \end{minipage}
    \end{minipage}
    \subcaption*{JPEG\\23.24 / 0.1833}
  \end{minipage}
  \begin{minipage}{0.32\hsize}
    \centering
    \begin{minipage}{0.73\hsize}
      \centering
      \includegraphics[width=\hsize]{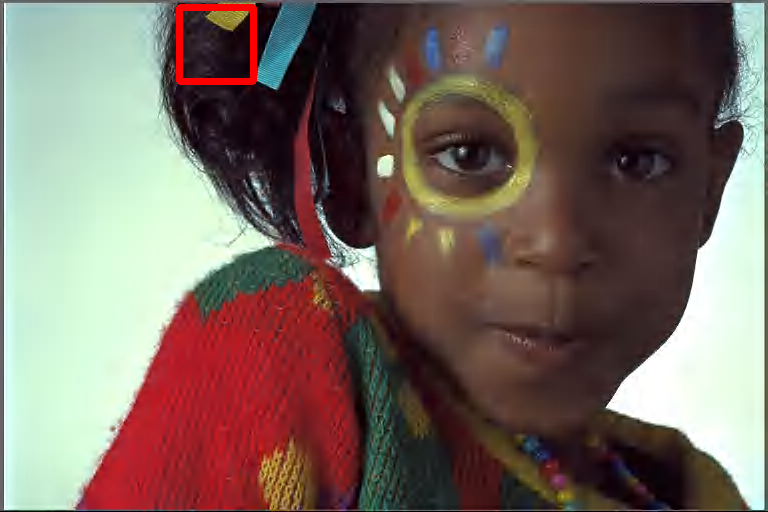}
    \end{minipage}
    \begin{minipage}{0.25\hsize}
      \begin{minipage}{0.9\hsize}
        \centering
        \setlength{\fboxrule}{1pt}
        \setlength{\fboxsep}{0pt}
        \fcolorbox{red}{white}{\includegraphics[width=\hsize]{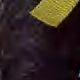}}
      \end{minipage}\\
      \begin{minipage}{0.9\hsize}
      \end{minipage}\\
      \begin{minipage}{0.9\hsize}
        \centering
        \setlength{\fboxrule}{1pt}
        \setlength{\fboxsep}{0pt}
        \fcolorbox{red}{white}{\includegraphics[width=\hsize]{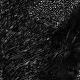}}
      \end{minipage}
    \end{minipage}
    \subcaption*{JPEG2000\\30.05 / 0.1778}
  \end{minipage}\\
  \begin{minipage}{0.32\hsize}
    \centering
    \begin{minipage}{0.73\hsize}
      \centering
      \includegraphics[width=\hsize]{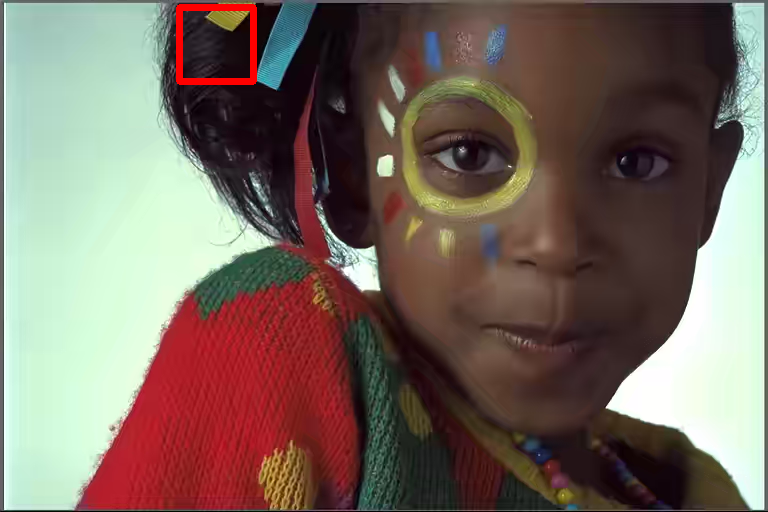}
    \end{minipage}
    \begin{minipage}{0.25\hsize}
      \begin{minipage}{0.9\hsize}
        \centering
        \setlength{\fboxrule}{1pt}
        \setlength{\fboxsep}{0pt}
        \fcolorbox{red}{white}{\includegraphics[width=\hsize]{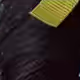}}
      \end{minipage}\\
      \begin{minipage}{0.9\hsize}
      \end{minipage}\\
      \begin{minipage}{0.9\hsize}
        \centering
        \setlength{\fboxrule}{1pt}
        \setlength{\fboxsep}{0pt}
        \fcolorbox{red}{white}{\includegraphics[width=\hsize]{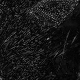}}
      \end{minipage}
    \end{minipage}
    \subcaption*{BPG\\31.81 / 0.1784}
  \end{minipage}
  \begin{minipage}{0.32\hsize}
    \centering
    \begin{minipage}{0.73\hsize}
      \centering
      \includegraphics[width=\hsize]{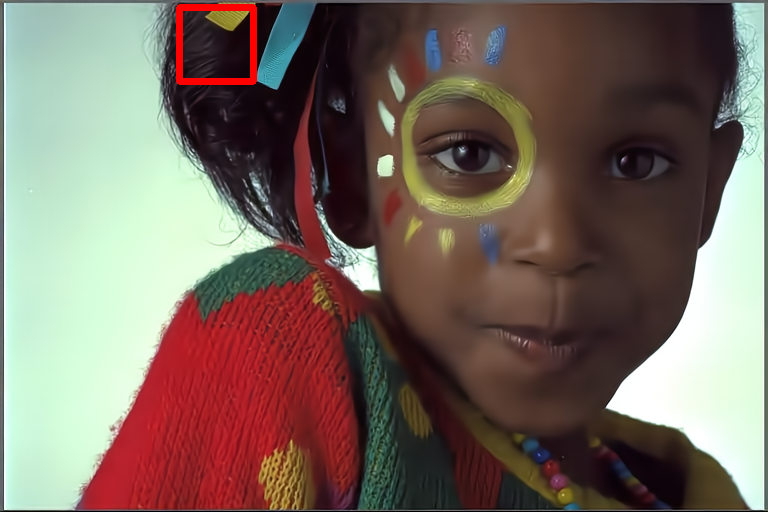}
    \end{minipage}
    \begin{minipage}{0.25\hsize}
      \begin{minipage}{0.9\hsize}
        \centering
        \setlength{\fboxrule}{1pt}
        \setlength{\fboxsep}{0pt}
        \fcolorbox{red}{white}{\includegraphics[width=\hsize]{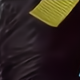}}
      \end{minipage}\\
      \begin{minipage}{0.9\hsize}
      \end{minipage}\\
      \begin{minipage}{0.9\hsize}
        \centering
        \setlength{\fboxrule}{1pt}
        \setlength{\fboxsep}{0pt}
        \fcolorbox{red}{white}{\includegraphics[width=\hsize]{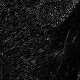}}
      \end{minipage}
    \end{minipage}
    \captionsetup{justification=centering}
    \subcaption*{Cheng20~\cite{conf/cvpr/Cheng2020} (AUN-Q)\\32.70 / 0.1775}
  \end{minipage}
  \begin{minipage}{0.32\hsize}
    \centering
    \begin{minipage}{0.73\hsize}
      \centering
      \includegraphics[width=\hsize]{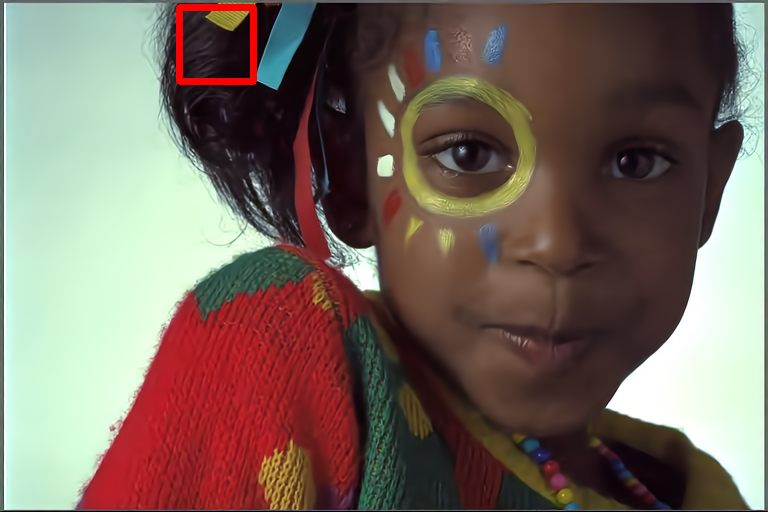}
    \end{minipage}
    \begin{minipage}{0.25\hsize}
      \begin{minipage}{0.9\hsize}
        \centering
        \setlength{\fboxrule}{1pt}
        \setlength{\fboxsep}{0pt}
        \fcolorbox{red}{white}{\includegraphics[width=\hsize]{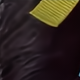}}
      \end{minipage}\\
      \begin{minipage}{0.9\hsize}
      \end{minipage}\\
      \begin{minipage}{0.9\hsize}
        \centering
        \setlength{\fboxrule}{1pt}
        \setlength{\fboxsep}{0pt}
        \fcolorbox{red}{white}{\includegraphics[width=\hsize]{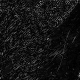}}
      \end{minipage}
    \end{minipage}
    \captionsetup{justification=centering}
    \subcaption*{Cheng20 + ours (AUN-Q \& U-Q)\\32.81 / 0.1727}
  \end{minipage}
  \caption{Qualitative results on the Kodak dataset. The bottom right image denotes the absolute difference magnified by four between the original and the compressed image.}
  \label{fig:qualitative}
\end{figure*}

\begin{figure*}[t]
  \begin{minipage}{0.32\hsize}
    \centering
    \begin{minipage}{0.73\hsize}
      \centering
      \includegraphics[width=\hsize]{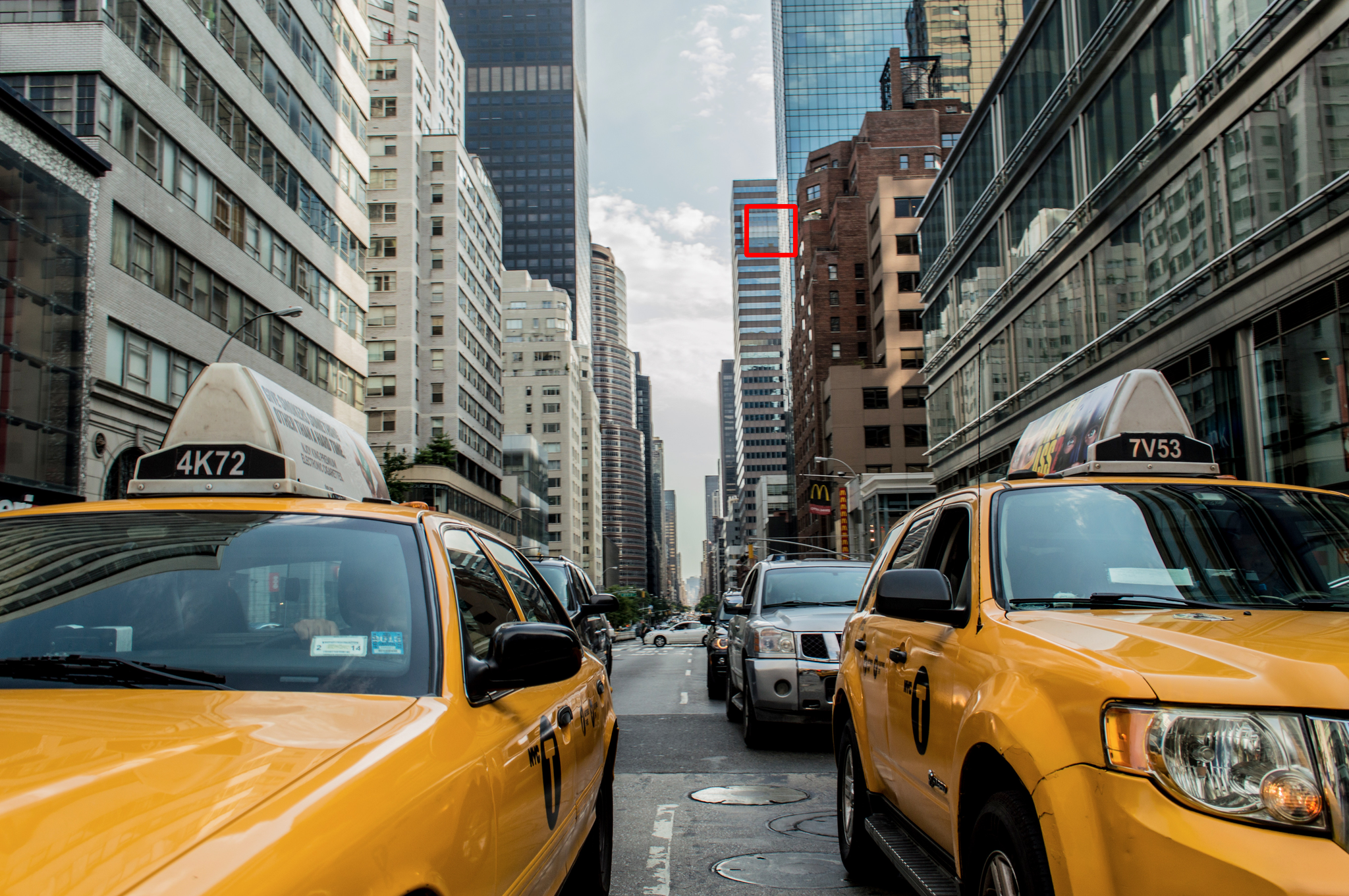}
    \end{minipage}
    \begin{minipage}{0.25\hsize}
      \begin{minipage}{0.9\hsize}
        \centering
        \setlength{\fboxrule}{1pt}
        \setlength{\fboxsep}{0pt}
        \fcolorbox{red}{white}{\includegraphics[width=\hsize]{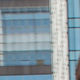}}
      \end{minipage}\\
      \begin{minipage}{0.9\hsize}
      \end{minipage}\\
      \begin{minipage}{0.9\hsize}
        \centering
        \setlength{\fboxrule}{1pt}
        \setlength{\fboxsep}{0pt}
        \fcolorbox{red}{white}{\includegraphics[width=\hsize]{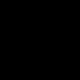}}
      \end{minipage}
    \end{minipage}
    \captionsetup{justification=centering}
    \subcaption*{Original Image\\PSNR / BPP}
  \end{minipage}
  \begin{minipage}{0.32\hsize}
    \centering
    \begin{minipage}{0.73\hsize}
      \centering
      \includegraphics[width=\hsize]{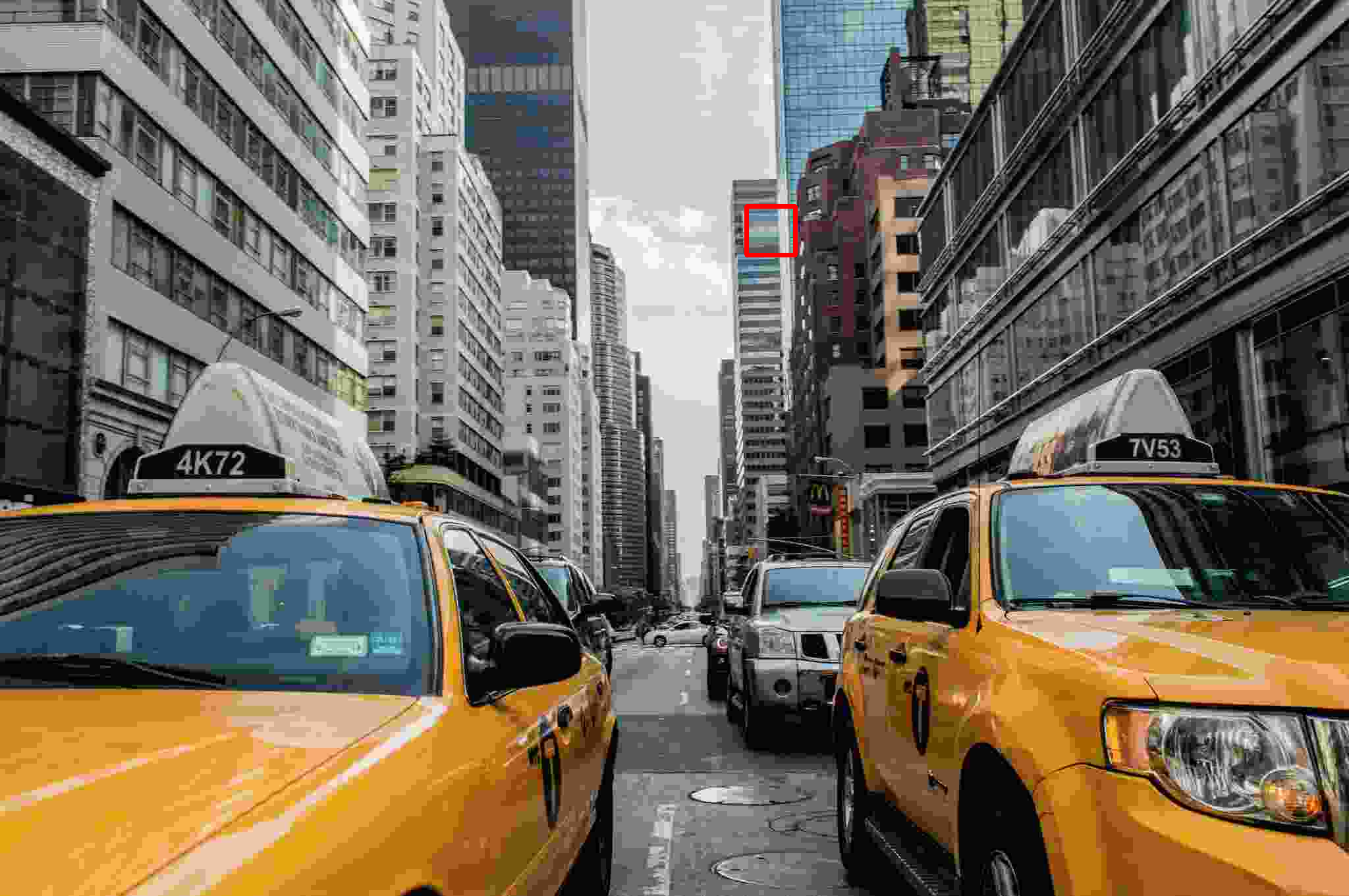}
    \end{minipage}
    \begin{minipage}{0.25\hsize}
      \begin{minipage}{0.9\hsize}
        \centering
        \setlength{\fboxrule}{1pt}
        \setlength{\fboxsep}{0pt}
        \fcolorbox{red}{white}{\includegraphics[width=\hsize]{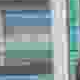}}
      \end{minipage}\\
      \begin{minipage}{0.9\hsize}
      \end{minipage}\\
      \begin{minipage}{0.9\hsize}
        \centering
        \setlength{\fboxrule}{1pt}
        \setlength{\fboxsep}{0pt}
        \fcolorbox{red}{white}{\includegraphics[width=\hsize]{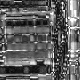}}
      \end{minipage}
    \end{minipage}
    \captionsetup{justification=centering}
    \subcaption*{JPEG\\24.87 / 0.2267}
  \end{minipage}
  \begin{minipage}{0.32\hsize}
    \centering
    \begin{minipage}{0.73\hsize}
      \centering
      \includegraphics[width=\hsize]{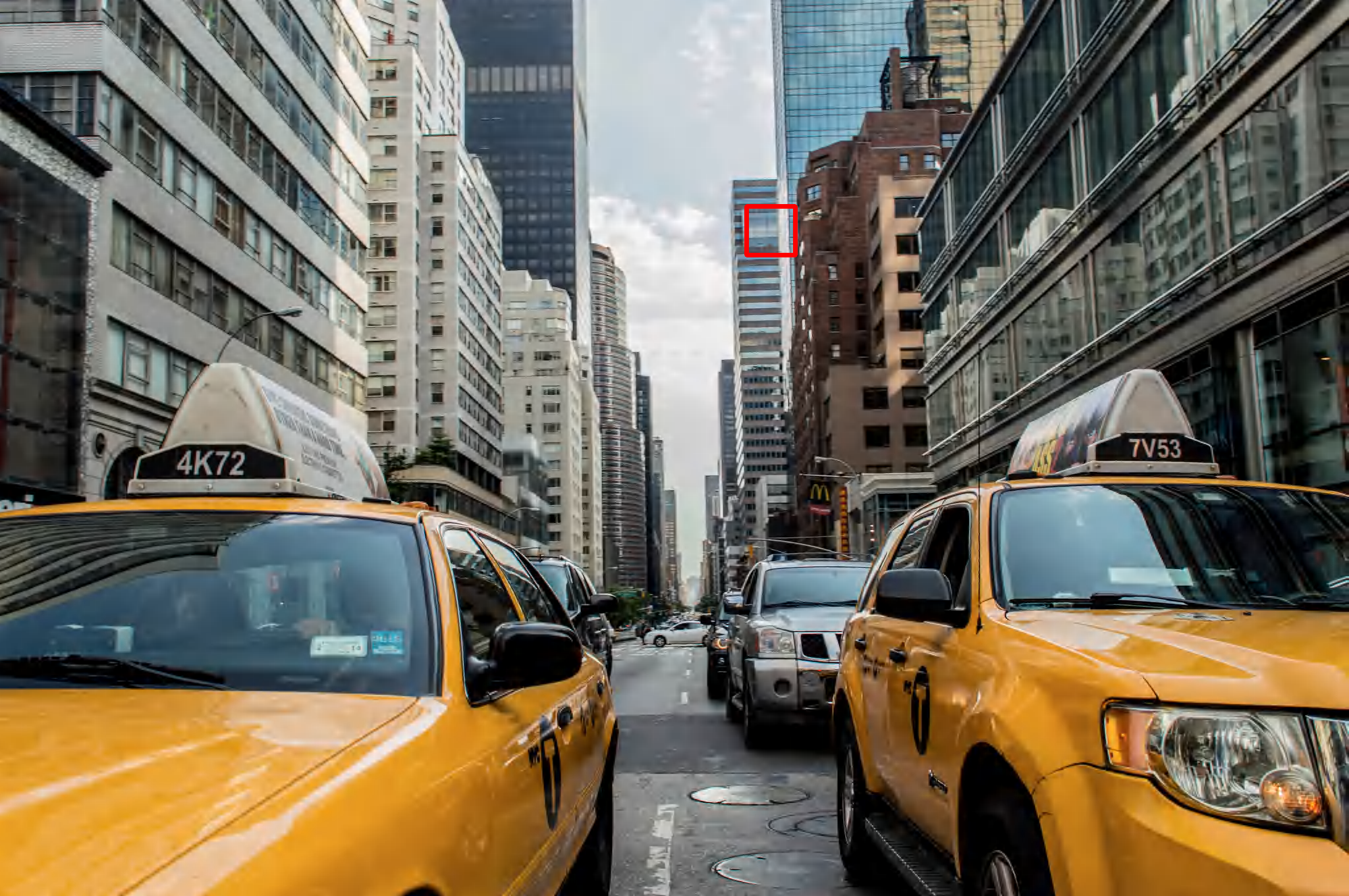}
    \end{minipage}
    \begin{minipage}{0.25\hsize}
      \begin{minipage}{0.9\hsize}
        \centering
        \setlength{\fboxrule}{1pt}
        \setlength{\fboxsep}{0pt}
        \fcolorbox{red}{white}{\includegraphics[width=\hsize]{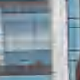}}
      \end{minipage}\\
      \begin{minipage}{0.9\hsize}
      \end{minipage}\\
      \begin{minipage}{0.9\hsize}
        \centering
        \setlength{\fboxrule}{1pt}
        \setlength{\fboxsep}{0pt}
        \fcolorbox{red}{white}{\includegraphics[width=\hsize]{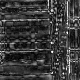}}
      \end{minipage}
    \end{minipage}
    \captionsetup{justification=centering}
    \subcaption*{JPEG2000\\29.57 / 0.2242}
  \end{minipage}\\
  \begin{minipage}{0.32\hsize}
    \centering
    \begin{minipage}{0.73\hsize}
      \centering
      \includegraphics[width=\hsize]{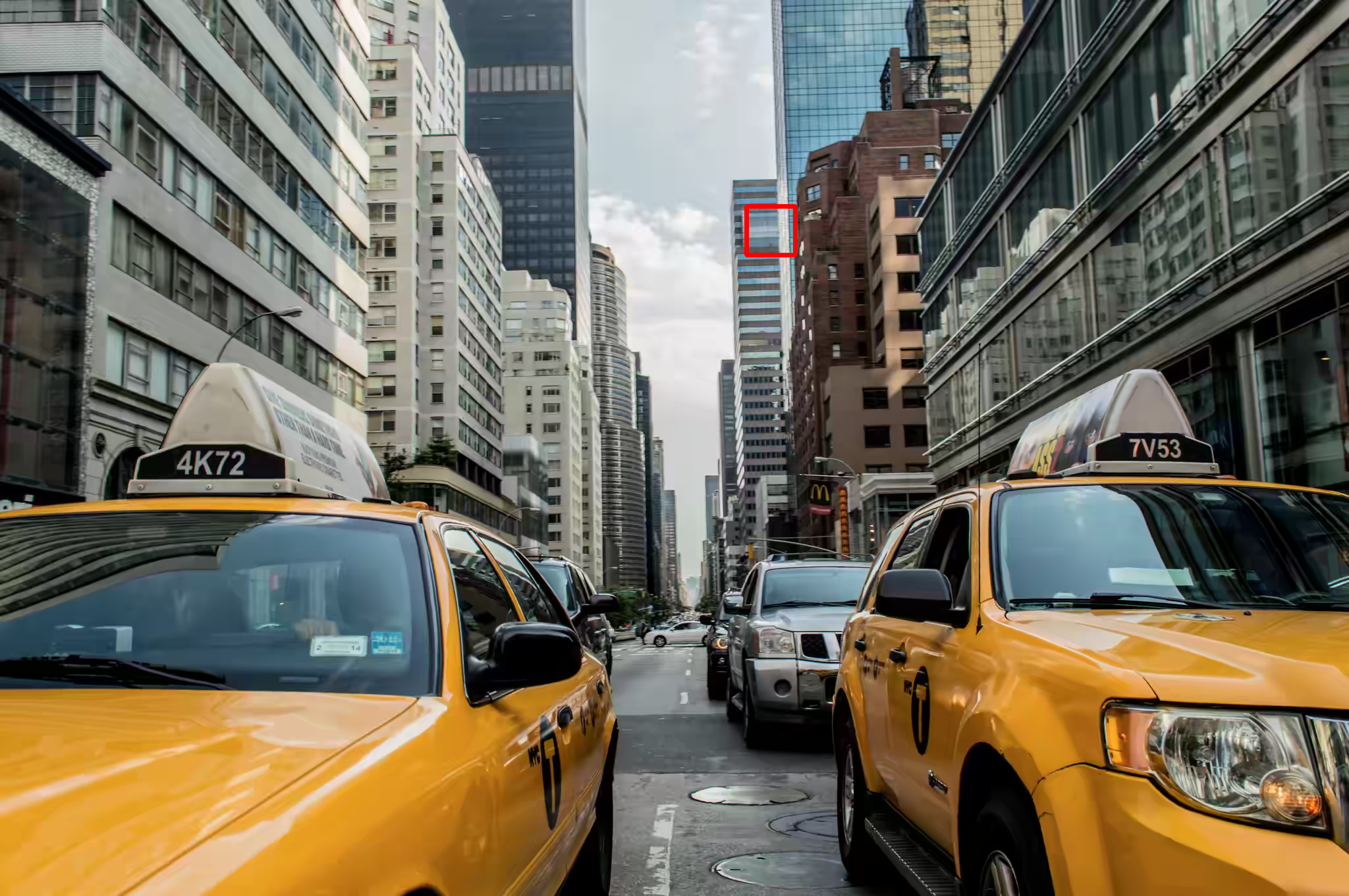}
    \end{minipage}
    \begin{minipage}{0.25\hsize}
      \begin{minipage}{0.9\hsize}
        \centering
        \setlength{\fboxrule}{1pt}
        \setlength{\fboxsep}{0pt}
        \fcolorbox{red}{white}{\includegraphics[width=\hsize]{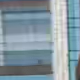}}
      \end{minipage}\\
      \begin{minipage}{0.9\hsize}
      \end{minipage}\\
      \begin{minipage}{0.9\hsize}
        \centering
        \setlength{\fboxrule}{1pt}
        \setlength{\fboxsep}{0pt}
        \fcolorbox{red}{white}{\includegraphics[width=\hsize]{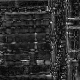}}
      \end{minipage}
    \end{minipage}
    \captionsetup{justification=centering}
    \subcaption*{BPG\\31.83 / 0.2213}
  \end{minipage}
  \begin{minipage}{0.32\hsize}
    \centering
    \begin{minipage}{0.73\hsize}
      \centering
      \includegraphics[width=\hsize]{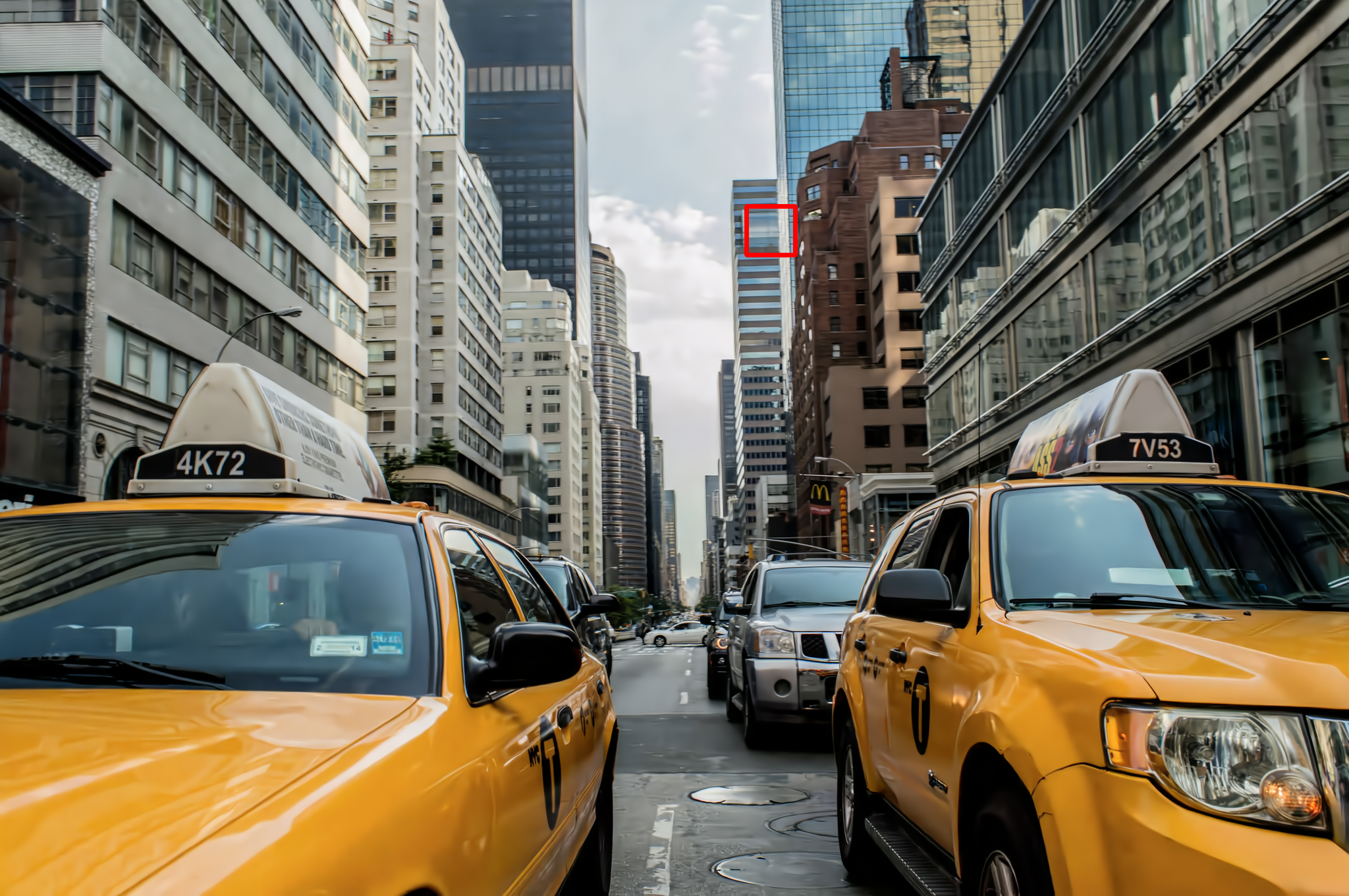}
    \end{minipage}
    \begin{minipage}{0.25\hsize}
      \begin{minipage}{0.9\hsize}
        \centering
        \setlength{\fboxrule}{1pt}
        \setlength{\fboxsep}{0pt}
        \fcolorbox{red}{white}{\includegraphics[width=\hsize]{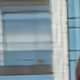}}
      \end{minipage}\\
      \begin{minipage}{0.9\hsize}
      \end{minipage}\\
      \begin{minipage}{0.9\hsize}
        \centering
        \setlength{\fboxrule}{1pt}
        \setlength{\fboxsep}{0pt}
        \fcolorbox{red}{white}{\includegraphics[width=\hsize]{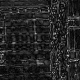}}
      \end{minipage}
    \end{minipage}
    \captionsetup{justification=centering}
    \subcaption*{Cheng20~\cite{conf/cvpr/Cheng2020} (AUN-Q)\\33.51 / 0.2233}
  \end{minipage}
  \begin{minipage}{0.32\hsize}
    \centering
    \begin{minipage}{0.73\hsize}
      \centering
      \includegraphics[width=\hsize]{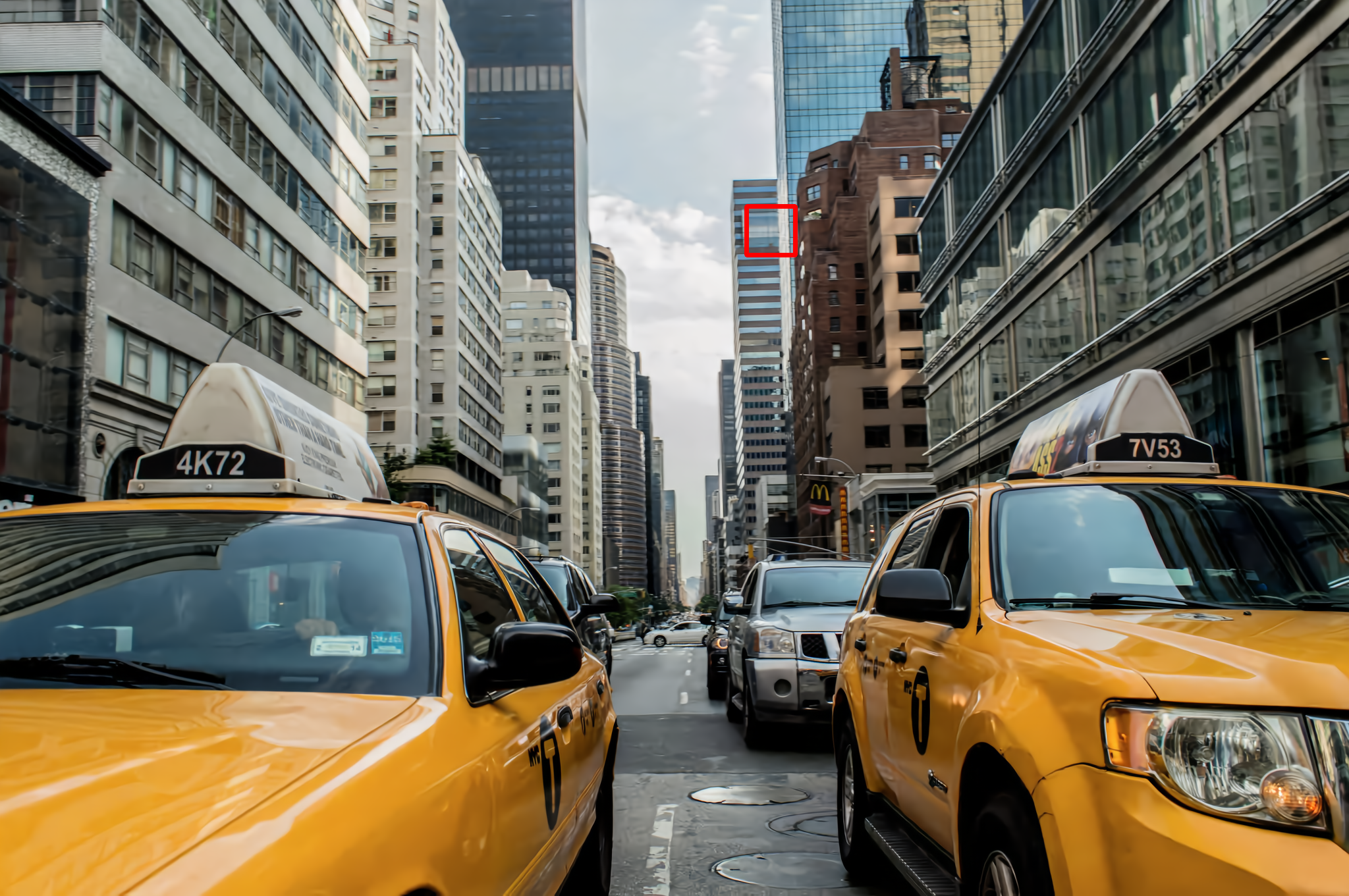}
    \end{minipage}
    \begin{minipage}{0.25\hsize}
      \begin{minipage}{0.9\hsize}
        \centering
        \setlength{\fboxrule}{1pt}
        \setlength{\fboxsep}{0pt}
        \fcolorbox{red}{white}{\includegraphics[width=\hsize]{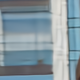}}
      \end{minipage}\\
      \begin{minipage}{0.9\hsize}
      \end{minipage}\\
      \begin{minipage}{0.9\hsize}
        \centering
        \setlength{\fboxrule}{1pt}
        \setlength{\fboxsep}{0pt}
        \fcolorbox{red}{white}{\includegraphics[width=\hsize]{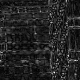}}
      \end{minipage}
    \end{minipage}
    \captionsetup{justification=centering}
    \subcaption*{Cheng20 + ours (AUN-Q \& U-Q)\\33.62 / 0.2170}
  \end{minipage}
  \caption{Qualitative results on the CLIC dataset. The bottom right image denotes the absolute difference magnified by seven between the original and the compressed image.}
  \label{fig:qualitative_clic}
\end{figure*}
We demonstrate the performance of the best approximation using the rate-distortion curve.
We used PSNR and BPP to evaluate the distortion and the rate, respectively, to plot the rate-distortion curve.
We used three network architectures (Ball\'{e}17, Ball\'{e}18, and Cheng20) that we used in the previous experiments, and adopt the best approximated quantization for each network architecture.
We compared them with these network architectures using original approximated quantization.
We also compared with traditional methods, \ie, JPEG~\cite{JPEG}, JPEG2000~\cite{JPEG2000}, and BPG~\cite{BPG}, for reference.
These methods are implemented in OpenCV, OpenJPEG~\cite{OpenJPEG}, and \cite{BPG}.
We ran these programs with the default configurations.
The results on the Kodak dataset are shown in Fig.~\ref{fig:comparison}.
As stated in the previous experiments, using the best approximation is better than using the original approximation.

We present the qualitative results for the Kodak dataset in Fig.~\ref{fig:qualitative}, and the CLIC dataset in Fig.~\ref{fig:qualitative_clic}, which show the results of three traditional methods, Cheng20, and Cheng20 using the best approximation (Cheng20 + ours).
The results demonstrate that Cheng20 + ours achieves better visual quality with a lower bitrate than traditional methods.
Compared to Cheng20, we can observe an improvement in visual quality.
For example, in the upper right of the cropped patches in Fig.~\ref{fig:qualitative}, we can see clear stripes in our method.
In the center of the cropped patches in Fig.~\ref{fig:qualitative_clic}, we can see relatively clear grid lines in Cheng20 + ours.

\begin{table}[t]
  \centering
  \caption{Per-patch evaluation of test images by difficulty levels. MSE denotes the mean squared error and BPP denotes bits per pixel.}
  \label{tbl:analysis}
  \begin{tabular}{@{}lcccccc@{}}
    \toprule
    & \multicolumn{2}{c}{Easy} & \multicolumn{2}{c}{Medium} & \multicolumn{2}{c}{Hard}\\
    \cmidrule(lr){2-3}\cmidrule(lr){4-5}\cmidrule(l){6-7}
    & MSE$\downarrow$ & BPP$\downarrow$ & MSE$\downarrow$ & BPP$\downarrow$ & MSE$\downarrow$ & BPP$\downarrow$\\
    \midrule
    Cheng20        & 7.0 & 0.031 & 39.2 & 0.216 & 101.8 & \textbf{0.728}\\
    Cheng20 + ours & \textbf{6.7} & \textbf{0.028} & \textbf{39.0} & \textbf{0.200} & \textbf{98.3}  & 0.735\\
    \midrule
    \midrule
    Improvement & \textbf{4.3\%} & \textbf{10.1\%} & \textbf{0.5\%} & \textbf{7.7\%} & \textbf{3.4\%} & -0.9\%\\
    \bottomrule
  \end{tabular}
\end{table}

\begin{figure}[t]
  \begin{minipage}{\hsize}
    \centering
    \includegraphics[width=0.95\hsize]{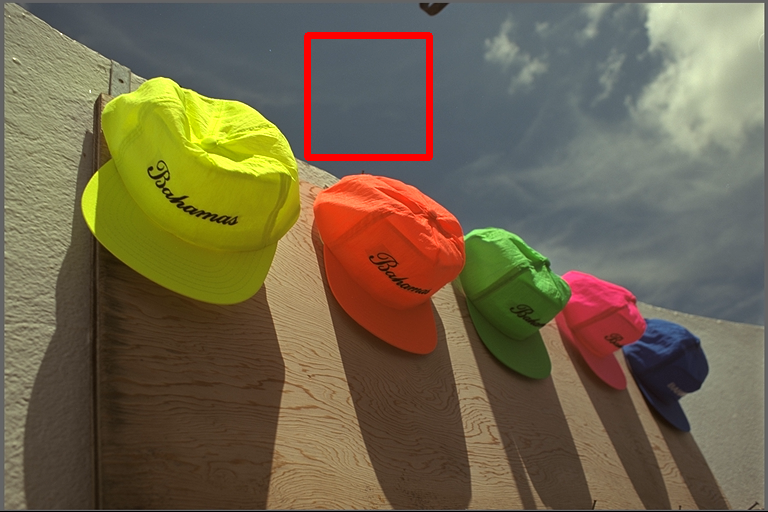}
  \end{minipage}\\
  \begin{minipage}{0.32\hsize}
    \centering
    \setlength{\fboxrule}{1pt}
    \setlength{\fboxsep}{0pt}
    \fcolorbox{red}{white}{\includegraphics[width=\hsize]{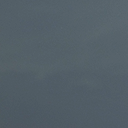}}
    \captionsetup{justification=centering}
    \subcaption*{Original Image}
  \end{minipage}
  \begin{minipage}{0.32\hsize}
    \centering
    \setlength{\fboxrule}{1pt}
    \setlength{\fboxsep}{0pt}
    \fcolorbox{red}{white}{\includegraphics[width=\hsize]{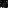}}
    \captionsetup{justification=centering}
    \subcaption*{Cheng20 (AUN-Q)}
  \end{minipage}
  \begin{minipage}{0.32\hsize}
    \centering
    \setlength{\fboxrule}{1pt}
    \setlength{\fboxsep}{0pt}
    \fcolorbox{red}{white}{\includegraphics[width=\hsize]{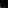}}
    \captionsetup{justification=centering}
    \subcaption*{Cheng20 + ours}
  \end{minipage}
  \caption{Visualization of the bit allocation. We can observe the bit reduction on the flat region.}
  \label{fig:analysis}
\end{figure}

We investigated why the best approximation improves the performance by analyzing per patch.
We conducted this analysis on Cheng20 with $\lambda = 0.0075$ using the Kodak dataset.
We split each test image into $16 \times 16$ patches and classified them into three categories based on the difficulty levels of compression: easy, medium, and hard.
We defined the difficulty levels by the test loss of Cheng20 and treated patches in the best 25\% as easy patches, patches in the worst 25\% as hard patches, and the other patches as medium patches.
To compute the loss function, we measured the bitrate of each patch based on the log-likelihood of the corresponding pixel in the latent representation.
We show the performance by difficulty levels in Table~\ref{tbl:analysis}.
The results indicate the best approximation improves the performance on easier patches.
We also investigated the reason and found that the number of bits reduces on the flat region as shown in Fig.~\ref{fig:analysis}.
 
\subsection{Comparison Between Uniform and Non-Uniform Quantization}
In this section, we compare uniform and non-uniform quantization.
We compare AUN-Q, a standard uniform quantization, with a basic non-uniform quantization method (NU-Q) proposed in \cite{conf/cvpr/MentzerATTG18}.
NU-Q quantizes the input values into learnable centers whereas AUN-Q quantizes the input values into a fixed integer grid.

It is challenging to compare these quantization methods in the same condition because the architecture of entropy models for these methods is completely different.
Therefore, we adopted an advanced entropy model called a 3D-CNN-based model~\cite{conf/cvpr/MentzerATTG18} for NU-Q and a primitive entropy model called a factorized-prior model~\cite{conf/iclr/BalleLS17} for AUN-Q.
This makes it possible to state that AUN-Q is superior to NU-Q when AUN-Q \& factorized-prior model performs superior to NU-Q \& 3D-CNN-based model.
Note that the 3D-CNN-based model is more advanced because it uses adjacent pixels and channels as context, unlike the factorized-prior model.

\begin{table}[t]
  \centering
  \caption{Comparison between uniform quantization (AUN-Q~\cite{conf/iclr/BalleLS17}) and non-uniform quantization (NU-Q~\cite{conf/cvpr/MentzerATTG18}) in BD rate (\%). A smaller value is more effective.}
  \label{tbl:nonuq}
  \begin{tabular}{cccc}
    \toprule
    & \multicolumn{3}{c}{Encoder / Decoder}\\
    \cmidrule(lr){2-4}
    Quantization & Ball\'{e}17 & Ball\'{e}18 & Cheng20\\
    \midrule
    Uniform (AUN-Q~\cite{conf/iclr/BalleLS17}) & \textbf{0.00} & \textbf{0.00} & \textbf{0.00}\\
    Non-Uniform (NU-Q~\cite{conf/cvpr/MentzerATTG18}) & 17.08 & 18.15 & 14.92\\
    \bottomrule
  \end{tabular}
\end{table}

We conduct experiments based on the three architectures of the encoder and decoder: Ball\'{e}17, Ball\'{e}18, and Cheng20.
We modified the architectures of the encoder following the non-uniform quantization method~\cite{conf/cvpr/MentzerATTG18} -- we masked the latent representation with a binary mask obtained by increasing the number of output channels of the encoder.
We also modified the loss function for the bitrate following \cite{conf/cvpr/MentzerATTG18} -- we computed both the original and masked coding cost.
We set $\lambda$ to $\{0.001, 0.003, 0.01, 0.03\}$ for AUN-Q, and $\{0.0003, 0.001, 0.003, 0.01\}$ for NU-Q to align the BPP with AUN-Q.
We trained the models with these hyper-parameters once and computed the BD rate to AUN-Q.
To evaluate the bitrate, we used estimated BPP instead of actual BPP for ease of implementation.
Other settings are the same with Sec.~\ref{subsec:exp_setup}.

We show the experimental results on the Kodak dataset in Table~\ref{tbl:nonuq}.
Although an advanced entropy model is used for NU-Q, AUN-Q performs superior to NU-Q on all three architectures of the encoder and decoder.
This indicates that the performance difference is due to the type of quantization and that uniform quantization is superior to non-uniform quantization.
Comparison using the same entropy model leads to a more fair comparison and is a future work of our study.

\section{Conclusion}\label{sec:conclusion}
We performed comprehensive comparisons of approximated quantization for deep image compression.
We compared seven approximation methods and their combinations for a decoder and an entropy model.
We evaluate these methods using three standard network architectures on two datasets.
The experimental results demonstrate that the best approximation method outperforms existing approximation methods, whereas different approximation is best depending on the network architectures.
We also find that the combination of U-Q for the entropy model and DS-Q for the decoder is a good choice on average.

\bibliographystyle{IEEEtran}
\bibliography{refs}

\begin{IEEEbiography}[{\includegraphics[width=1in,height=1.25in,clip,keepaspectratio]{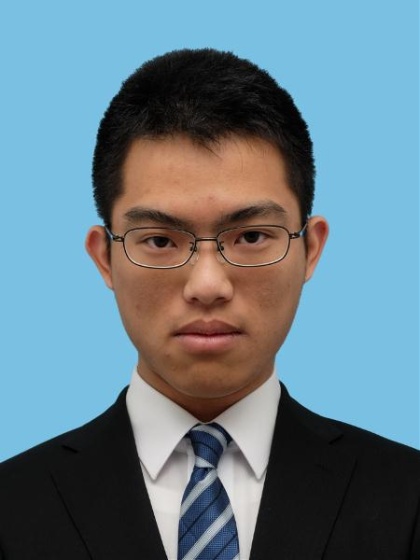}}]{Koki Tsubota} received the B.E. and M.S. from the University of Tokyo, in 2018, 2020, respectively. He is now a Ph.D. student of Department of Information and Communication Engineering. He is interested in computer vision and multimedia. He works on embedding learning, image compression, image quality assessment, and various topics of manga image processing -- manga object detection, manga character clustering, etc.
\end{IEEEbiography}

\begin{IEEEbiography}[{\includegraphics[width=1in,height=1.25in,clip,keepaspectratio]{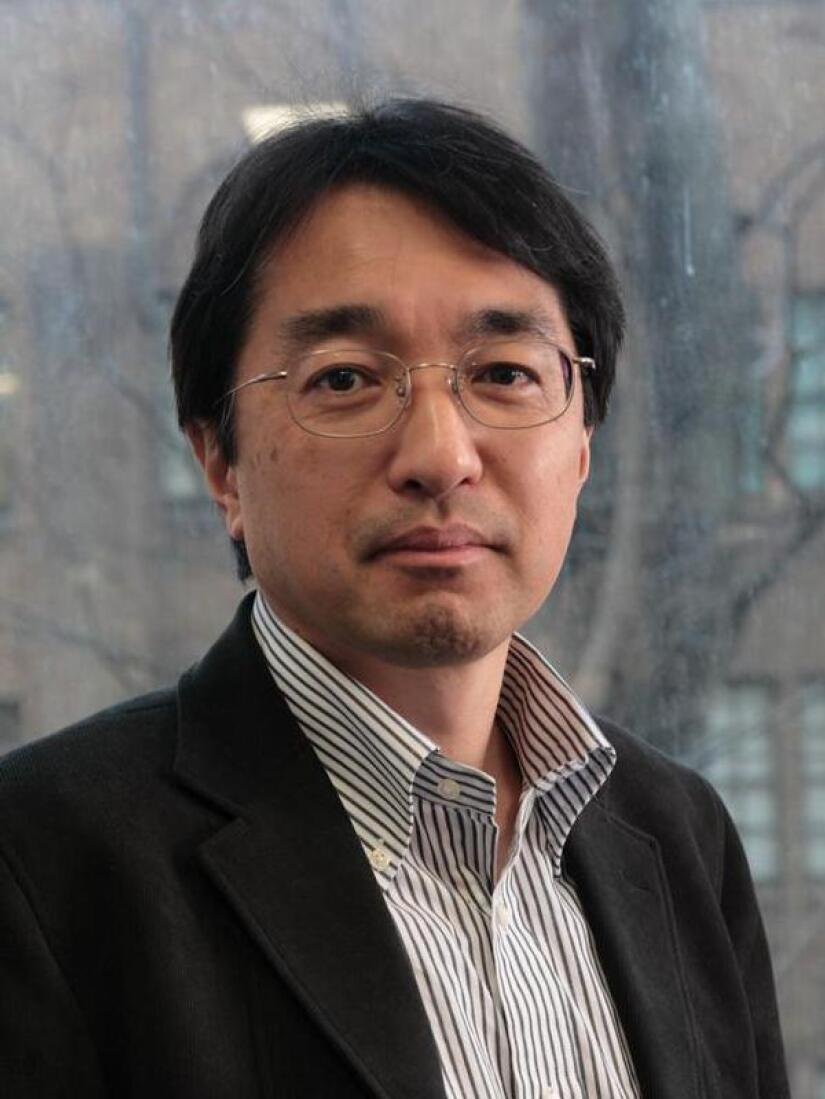}}]{Kiyoharu Aizawa} received the B.E., the M.E., and the Dr.Eng. degrees in Electrical Engineering all from the University of Tokyo, in 1983, 1985, 1988, respectively. He is currently a Professor at Department of Information and Communication Engineering of the University of Tokyo. He was a Visiting Assistant Professor at University of Illinois from 1990 to 1992. His research interest is in image processing and multimedia applications. He received the 1987 Young Engineer Award and the 1990, 1998 Best Paper Awards, the 1991 Achievement Award, 1999 Electronics Society Award from IEICE Japan, and the 1998 Fujio Frontier Award, the 2002 and 2009 Best Paper Award, and 2013 Achievement award from ITE Japan. He received the IBM Japan Science Prize in 2002. He is currently a Senior Associate Editor of IEEE Tras. Image Processing, and on Editorial Board of ACM TOMM, APSIPA Transactions on Signal and Information Processing, and International Journal of Multimedia Information Retrieval. He served as the Editor in Chief of Journal of ITE Japan, an Associate Editor of IEEE Trans. Image Processing, IEEE Trans. CSVT and IEEE Trans. Multimedia. He has served a number of international and domestic conferences; he was a General co-Chair of ACM Multimedia 2012. He is a council member of Science Council of Japan.
\end{IEEEbiography}

\EOD

\end{document}